\documentclass[preprint, 5p, twocolumn]{elsarticle}
\usepackage{amsmath}
\usepackage{amsfonts}
\usepackage{amssymb}
\usepackage{amsthm}

\usepackage{booktabs}
\usepackage{rotating}
\usepackage{tabu}

\usepackage{subcaption}

\usepackage{url}
\usepackage{algpseudocode}

\theoremstyle{definition}
\newtheorem{definition}{Definition}[section]

\journal{Journal of Network and Computer Applications}

\begin{document}

\begin{frontmatter}

\title{A Framework for Automating Deployment and Evaluation of \\Blockchain Networks}

\author[UofA]{Nguyen Khoi Tran}
\ead{nguyen.tran@adelaide.edu.au}

\author[UofA,CSCRC]{M. Ali Babar}

\author[DSTG]{Andrew Walters}

\affiliation[UofA]{organization={The University of Adelaide},
state={South Australia},
country={Australia}}

\affiliation[CSCRC]{organization={Cyber Security Cooperative Research Centre},
country={Australia}}

\affiliation[DSTG]{organization={Defence Science and Technology Group},
country={Australia}}


\begin{abstract}
    A blockchain network is a distributed system established by mutually distrusting participants to operate a blockchain, enabling them to manage critical information such as account balances or asset ownership without a centralised third party. Blockchain network deployment and evaluation have become prevalent due to the emerging blockchain use cases by enterprises, governments, and Internet of Things (IoT) applications, which demand private blockchains rather than participating in public ones. A blockchain network architecture drives deployment and evaluation activities. Nevertheless, practitioners must learn and perform error-prone activities to transform architecture into a blockchain network and evaluate it. Therefore, it is beneficial to automate these activities so that practitioners can focus on the architecture design, a valuable and hard-to-automate activity. The key challenges of such an automation framework are keeping up with the advances in blockchain technologies and the increasing complexity of blockchain network architecture. This paper proposes NVAL, a software framework that implements a novel architecture-driven, community-supported approach to automate blockchain network deployment and evaluation. NVAL accepts blockchain network architecture as input. It supports complex multi-channel blockchain networks, an increasingly prevalent architecture for private blockchain. The framework keeps up with blockchain technologies by leveraging platform-specific automation programs developed by a practitioner community via runtime composition to handle new networks. We evaluated NVAL with a case study and showed that the framework requires only seven automation programs to deploy 65 blockchain networks with 12 diverse architectures and generate 295 evaluation datasets. Furthermore, it consumes only 95.5 ms to plan and orchestrate the deployment and evaluation, which is minuscule compared to the total time required for deploying and benchmarking a blockchain network.
\end{abstract}



\begin{keyword}
Blockchain \sep Distributed Ledger \sep Deployment \sep Evaluation \sep Automation \sep Framework
\end{keyword}

\end{frontmatter}


\section{Introduction}
\label{sec:Introduction}



A \textit{blockchain} can be defined as a secure database and a replicated state machine operated by a distributed network of mutually distrusting participants \cite{Androulaki2018,Tran2021a,wood2014ethereum}. It helps participants to maintain and update shared state information, such as account balances or asset ownership, \textit{without relying on a centralised third party} \cite{nakamoto2008bitcoin}. The database contains \textit{transactions} that describe state updates such as transferring currency between accounts. The state machine processes transactions to calculate a new state, such as new account balances after currency transfer \cite{wood2014ethereum}. Both blockchain's database and state machine are replicated across participants and synchronised by a fault tolerant consensus protocol \cite{cachin2017blockchain}, ensuring that participants can access the state information and detect divergence by others. By embedding data and logic into blockchain's transactions, state variables, and state machines, participants conduct business processes without appointing a third party, reducing the failure risks and insider threats \cite{Buterin2017,xu2016blockchain}. Blockchains have been leveraged successfully for financial use cases such as cryptocurrency (digital cash) \cite{swan2015blockchain}, digital assets \cite{kugler2021non}, and decentralised finance platforms \cite{Avgouleas2020}. Blockchains have also been used as a secure timestamping service to prove the existence and authenticity of documents and a means for resisting information censorship \cite{swan2015blockchain}. Most of the described use cases utilise global \textit{public blockchains} such as Ethereum and Bitcoin. However, there are emerging use cases by enterprises, governments, edge computing, and the Internet of Things (IoT) \cite{Tran2021,Pawczuk2020} where public blockchains fail to provide the necessary performance, scalability, and confidentiality \cite{Xu2017}. Therefore, there has been an increasing demand for \textit{permissioned or private blockchains}. For instance, when conducting the first blockchain-based Letter-of-Credit (LC) transaction between Vietnam and Korea on July 3, 2019, HSBC utilised a private blockchain based on R3's Corda platform \cite{brown2018corda} rather than a public one.

A distributed system established by mutually distrusting participants to run a blockchain is known as a \textit{blockchain network} \cite{Tran2020}. It is made up of distributed computers running a \textit{blockchain client}, which is a software implementation of a blockchain protocol. Public blockchains run on public blockchain networks that emerge without coordination anonymous Internet users. Private blockchains are operated by private blockchain networks, which can be designed, deployed, and evaluated to ensure they fit an intended use case and hardware infrastructure \cite{Tran2020,Tran2021a}. Let us elaborate on the design, deployment, and evaluation of blockchain networks with the following scenario. 

\textbf{Motivating Scenario:} Multiple research labs from different universities wish to connect their robotic testbeds to form a joint experimentation platform to expand the scope of their experiments. For instance, researchers can use the platform to validate a robot coordination mechanism across multiple enviroments by assigning tasks to and receiving experiment data from edge agents such as uncrewed ground vehicles (UGV), uncrewed aerial vehicles (UAV), and uncrewed underwater vehicles (UUV) from other testbeds. A joint experimentation platform raises various security requirements: edge agents must verify the authenticity of incoming mission objectives \cite{Liang2017}, researchers must verify the integrity and provenance (history) of the experiment data \cite{Ramachandran2018}, and infrastructure owners must track researchers' requests for auditing. The information necessary for the above requirements constitutes the platform's state (e.g., tasking status of a robot) and the state updates (e.g., a researcher assigning a task to a robot). Participants can maintain such information securely without hiring a third party or appointing a leader to operate a centralised service by leveraging a private blockchain. Doing so requires establishing a private blockchain network. Participants can control all aspects of the network, such as where to deploy its constituing blockchain clients and how to configure them. Participants can even split the network into separate \textit{blockchain channels} that store different information and run different blockchain protocols. For instance, an inter-lab channel connecting powerful workstations can employ a secure but resource-intensive blockchain protocol whilst intra-lab channels between robots and base stations within a testbed can leverage a lightweight blockchain protocol (e.g., \cite{Bandara2021}). The described \textit{multi-channel heterogeneous blockchain network architecture} is increasingly prevalent in edge computing and IoT blockchain use cases \cite{Tran2021,Tran2021a} as it offers practitioners a lot of flexibility to fine-tune their blockchain network. 

A blockchain network architecture becomes an operational network via the \textit{blockchain network deployment} process, which involves provisioning and configuring blockchain clients according to the architecture \cite{Zheng2019}. Network deployment can be followed by \textit{blockchain network evaluation}, the assessment of the blockchain network's quality attributes such as latency and throughput via benchmarks \cite{Dinh2017}, or simulations \cite{Yasaweerasinghelage2017,Wang2018,Zhang2018,Aoki2019}. The faster participants iterate over deployment and evaluation, the more they can refine their architectures within limited time and resources \cite{bass2003software}. Unfortunately, blockchain network deployment and evaluation are time-consuming, error-prone, and dependent on platform-specific knowledge \cite{Lu2019,Malik2019}. Therefore, it is beneficial to improve the effectiveness and efficiency of blockchain network deployment and evaluation via automation. In fact, ``one-click deployment'' is a highly demanded tool for development and testing according to a recent survey on the state of practice of blockchain-oriented software engineering \cite{Bosu2019}.

\begin{figure*}[ht]
    \centering
    \includegraphics[width=0.8\textwidth,,trim={0cm 6cm 0cm 0cm},clip]{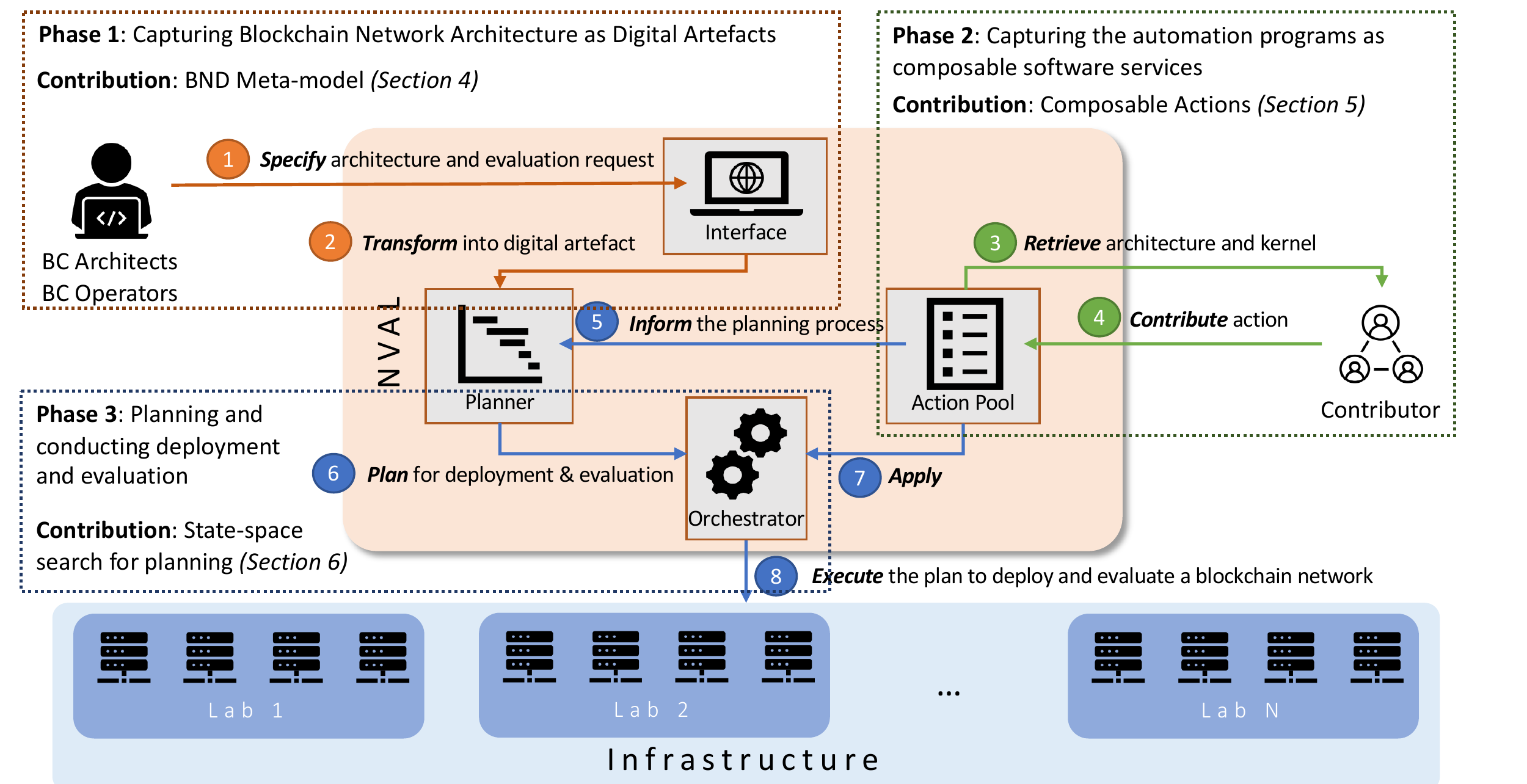}
    \caption{Users, Workflows, and Components of NVAL Framework}
    \label{fig:NVAL-overview}
\end{figure*}

Existing literature reports various approaches and tools for automating blockchain network deployment and evaluation. Practitioners can utilise generic IT automation tools such as Ansible\footnote{\url{https://www.ansible.com}} and Terraform\footnote{\url{https://www.terraform.io}} to perform repetitive activities such as provisioning and starting blockchain clients across computers to reduce time and implementation errors. However, generic tools fail to reduce complexity because practitioners must manually transform blockchain network architectures into tools' instructions. A recent study on DevOps practices in blockchain-based software engineering found little evidence of real-world utilisation of generic automation tools \cite{Woehrer2021}. Blockchain-specific automation frameworks such as PlaTIBART \cite{Walker2017}, Hyperledger Composer \cite{Hyperledger2019} and Hyperledger Bevel \cite{Hyperledger2021} (formerly known as the Blockchain Automation Framework) and MixBytes Tank \cite{MixBytes2021} have been proposed to address the limitation of generic automation tools. Blockchain-as-a-Service (BaaS) platforms such as uBaaS \cite{Lu2019}, NutBaaS \cite{Zheng2019}, AWS Managed Blockchain, Azure BaaS, and IBM Blockchain Platform are cloud services that also provide ``one-click deployment'' to provision blockchain networks on cloud-based virtual machines. While these tools and services are better than generic automation tools at abstracting the complexity, we found that most require low-level and technology-specific specifications (Compose, Bevel, PlaTIBART) or overly-simplified architectural models (BaaS platforms), making them too complex or restrictive to users. Moreover, we found that their applicability and longevity also suffer from the following limitations:
\begin{itemize}
    \item \textit{Lack of support for heterogeneous multi-channel blockchain networks}, a complex blockchain network architecture that is gaining popularity in blockchain use cases that involve edge computing and IoT \cite{Tran2021,Tran2021a}. The heterogeneity of this architecture stands in stark contrast to the lack of support for multi-channel topology and multi-technology-per-network of most current tools and services.  
    \item \textit{Limited scope of the supported blockchain platforms.} Blockchain platforms can be considered concrete implementations of blockchain protocols, providing software clients and utilities for deploying or participating in a blockchain network \cite{Tran2021}. We found that most of the current tools only support a few prominent blockchain platforms. Hyperledger Bevel, the state-of-the-art tool with the largest scope at the time of writing, supports only four blockchain platforms: Fabric, Corda, Indy, and Quorum. Meanwhile, a recent empirical study \cite{Das2022} of 3,664 GitHub repositories reveals 71 blockchain platforms. Should these platforms be supported by an automation tool, they would greatly expand the blockchain network design options for practitioners and help them establish more optimal blockchain networks for their use cases. Unfortunately, we believe that the current automation approach which targets specific blockchain platforms and relies on a core developer group to approve extension (e.g., via GitHub pull requests), might not scale and keep up with the advances in blockchain technology. In fact, one of the earlier automation tools for the Fabric blockchain platform, called Hyperledger Composer, was deprecated in 2021 as Fabric matures. 
\end{itemize}

This paper proposes \textit{NVAL (Network Deployment and Evaluation framework)}, a software framework that implements a novel architecture-driven, community-supported approach to automate the deployment and evaluation of blockchain networks. The framework is driven by blockchain network architecture specifications that describe high-level design decisions (e.g., network topology, protocol choice) and support heterogeneous multi-channel networks. Instead of focusing on a few blockchain platforms and relying on the open-source contribution for extension, NVAL leverages the existing platform-specific automation programs developed by a practitioner community by combining them at runtime according to the given blockchain network architecture for deployment and evaluation. This approach was motivated by many open-source automation scripts for blockchain networks. As of February 2022, we found 52 open-source GitHub repositories on automated blockchain network deployment using Ansible, 67 repositories using Terraform, and 35 repositories using other scripting languages. Most blockchain platforms also provide deployment instructions in their documentation, which can be transformed into automation scripts. By leveraging the automation programs emerging from the community, we aim to help NVAL keep up with the advances in blockchain technology that could outpace any team working alone on the framework. 

Let us elaborate on NVAL's operation and utility via the motivating scenario. Figure \ref{fig:NVAL-overview} presents an overview of the framework's users and workflow. NVAL benefits \textit{blockchain network architects} and \textit{operators}, who are members of the participating research labs appointed to design and deploy a private blockchain network. Architects gather requirements (e.g., available hardware, required latency and throughput, the required level of availability, threat models) and leverage the existing architectural knowledge (e.g., \cite{Tran2020,Tran2021a}) to develop potential blockchain network architectures. Operators deploy and evaluate the given architectures with the help of NVAL. In order to leverage the framework, operators deploy an NVAL instance on a workstation that can access all computers participating in a blockchain network. Operators start a workflow by submitting a request that contains a blockchain network architecture and a list of evaluation metrics for measuring (Step 1). NVAL parses and transforms the request into an actionable digital artefact (Step 2). The framework relies on an \textit{action pool} containing automation programs (\textit{actions}) contributed by contributors via open-source or inner-source channels (Step 3 and Step 4) to process the request. NVAL leverages its action pool (Step 5) to construct an \textit{execution plan} that describes a sequence and inputs of service calls to actions in the pool (Step 6). The framework applies actions according to the plan (Step 7) to deploy and evaluate a blockchain network (Step 8). NVAL returns URLs to access the deployed blockchain network and some datasets containing the required evaluation metrics following a successful execution. If NVAL cannot resolve a plan, it declares the request unsatisfiable.

NVAL realises the stated approach with three novel elements. We propose a novel domain-specific meta-model called \textit{Blockchain Network Design (BND)} to capture heterogeneous multi-channel blockchain network architectures. We propose the concept and architecture of \textit{composable actions} to encapsulate and combine the existing automation programs to process new blockchain network architectures specified using BND. Finally, we propose a \textit{planning approach} based on a state-space search algorithm to match individual blockchain channels from a given blockchain network architecture with composable actions and construct an execution plan for deployment and evaluation. We demonstrated and evaluated the feasibility of NVAL by conducting a case study on an architecture development project that utilised an NVAL proof-of-concept system to evaluate diverse blockchain network architectures empirically. The results show that NVAL successfully leverages seven automation programs to deploy 65 blockchain networks with 12 diverse architectures and generate 295 evaluation datasets. Furthermore, we found that NVAL requires only a tenth of a second (95.5 ms) to plan and orchestrate the deployment and evaluation. This figure is minuscule compared to the total time required for deploying and evaluating a blockchain network, which ranges from 7 to 14 minutes. The contributions of this paper are:

\begin{itemize}
    \item We conceptualise and design the architecture of NVAL, a software framework for automating blockchain network deployment and evaluation. NVAL supports multi-channel heterogeneous blockchain networks and provides user-driven, community-supported extensibility to address the limitations of the current approaches. 
    \item We propose Blockchain Network Design (BND), a meta-model for capturing blockchain network architecture as digital artefacts to guide the automated deployment.
    \item We conceptualise and design the architecture of Composable Actions, which encapsulates the existing automation programs into portable software components for runtime composition. 
    \item We propose an automated planning approach for matching blockchain network architectures with composable actions and constructing execution plans. 
    \item We demonstrated and evaluated NVAL via a case study, showing its applicability and efficiency.
\end{itemize}

\section{Background}
\label{sec:Background}

\subsection{Blockchain Network}

A blockchain is a share-write database and a secure computing engine operated by mutually-distrusting parties \cite{cachin2017blockchain,Greenspan2015}. Its operation is based on the consensus between participants rather than a central authority \cite{Buterin2017}. Every participant hosts and controls some \textit{blockchain nodes}, which are computers running a software implementation of a blockchain protocol. Blockchain nodes form a peer-to-peer system called a \textit{blockchain network} to operate and secure a blockchain. Interactions between blockchain nodes follow a \textit{blockchain consensus protocol} that dictates how they agree on the validity and order of blockchain transactions. The Proof-of-Work (PoW) protocol, also known as Nakamoto consensus \cite{nakamoto2008bitcoin}, is one of the most popular blockchain consensus protocols for public blockchain networks. This protocol allows anonymous Internet users to operate a blockchain securely and resists Sybil attacks, where a participant creates multiple fake identities to influence the consensus process. As PoW-based protocols tend to require significant computing resources, private blockchain networks generally rely on alternative consensus protocols such as Proof-of-Authority (PoA), Raft \cite{Aublin2013}, or Practical Byzantine Fault Tolerance (pBFT) \cite{Castro1999}. Cachin and Vukolic \cite{cachin2017blockchain} provides a survey of prominent consensus protocols.

Blockchain nodes within a blockchain network can assume different roles, which give them different degrees of autonomy and control over a blockchain and demand different amounts of computing resources \cite{Tran2020}:
\begin{itemize}
    \item \textit{Full Nodes} hold complete blockchain replicas. They can verify transactions and execute software scripts embedded in transactions, also known as smart contracts. 
    \item \textit{Mining Nodes} participate in a consensus protocol to process and append transactions. Mining nodes are usually full nodes, but not the other way around.
    \item \textit{Lightweight Nodes} hold only ledger metadata and must rely on trusted full nodes to verify transactions. They trade security and independence for reduced resource consumption.
    \item \textit{Remote Nodes} are end-user clients responsible for crafting and signing transactions on behalf of human or machine users. They are also known as wallets. 
    \item \textit{Network Interfaces} offer APIs for remote nodes to interact with a blockchain.
\end{itemize}

Blockchain networks can be segmented into \textit{channels} \cite{Tran2020}, which connect different subsets of blockchain nodes and maintains separate ledgers. Channels in the same network do not necessarily use the same blockchain consensus protocol. For instance, the motivating scenario in Section \ref{sec:Introduction} introduced a multi-channel heterogeneous blockchain network that contains an inter-lab channel running a resource intensive protocol and multiple intra-lab channels utilizing lightweight protocols.

\subsection{Design Decisions and Architecture of a Blockchain Network}
\label{subsec:design_decisions}

Private blockchain networks can be designed and deployed in a controlled manner to ensure their fitness for purpose. For instance, a blockchain network's architect can decide to split the network into channels to create a separation between different user groups or business processes. An architect can also decide how mining nodes are assigned to participants, reflecting participants' authority. By controlling the blockchain network's topology, configuration, and deployment, an architect can fine-tune quality attributes of the network such as performance, resource consumption, availability, and security \cite{Tran2020,Tran2021a}. 

Tran et al. defined a design space of blockchain network with 19 types of decisions \cite{Tran2021}, which can be organized into four types
\begin{itemize}
    \item \textit{Topological design decisions} capture the overall shape of a blockchain network. They include how a network is segmented into channels and the number of blockchain nodes per channel.
    \item \textit{Organizational design decisions} capture the assignment of blockchain nodes to participants. This assignment reflects participants' governance power and authorization level in the network.
    \item \textit{Consensus design decisions} capture the type and configuration of the consensus protocols used by blockchain channels in the network. 
    \item \textit{Deployment design decisions} capture the mapping between blockchain nodes and targeted hardware infrastructure.
\end{itemize}

The architecture of a blockchain network is the collection of its design decisions. Architecture specifications can capture blockchain network architectures as digital artefacts that computers can understand and implement. In this paper, we denote the language for writing such specifications as a meta-model of blockchain network architecture. 

\section{Approach}
\label{sec:Approach}

NVAL leverages a novel architecture-driven, community-supported approach to automate the deployment and evaluation of complex multi-channel blockchain networks. This approach consists of three phases presented in Figure \ref{fig:NVAL-overview}. 

The first phase is capturing the abstract and implicit blockchain network architectures as concrete artefacts that NVAL can understand and process. This phase requires a meta-model that defines concepts and relationships necessary for architects to describe blockchain network architectures. We found that the existing view models (4+1 \cite{Kruchten1995}) and notations (Unified Modelling Language (UML)) lack a standardised way to describe blockchain network architectures, leading to ambiguity. Moreover, the existing view models tend to scatter the design decisions across multiple views, increasing the complexity of capturing and understanding design decisions and could lead to inconsistencies. Therefore, we propose a novel domain-specific meta-model for capturing blockchain network architectures called \textit{Blockchain Network Design (BND)}. The meta-model defines standardised concepts, relations, and architectural structures for modelling blockchain network architectures based on an existing design space \cite{Tran2020}. Every blockchain network architecture specified using BND \textit{simultaneously} describes organisational, logical, and physical aspects of a blockchain network, thus mitigating the problem of scattering design decisions across multiple views. Section \ref{sec:BND} introduces BND in details.

The second phase is capturing the automation programs as composable software services that NVAL can combine to deploy and evaluate multi-channel heterogeneous blockchain networks. We denote the captured automation programs as \textit{composable actions}. Operators can develop actions from scratch or refactor them from existing inner-source or open-source automation scripts. Realising the composable action concept raises two challenges: (1) defining action scopes such that composition is possible and (2) minimising effort in developing composable actions. Section \ref{sec:Actions} introduces our architectural solution for composable actions.  

The third phase is planning and conducting the deployment and evaluation of incoming requests, which contain blockchain network architectures and lists of metrics to measure. We regard the action planning for a request as \textit{a search task on its implicit state-space} with the goal state being the successful deployment and evaluation of a requested blockchain network. This decision stems from the following observations:
\begin{itemize}
    \item \textit{Every request has a state}, reflecting whether (1) it has been verified, (2) its channels have been deployed, and (3) its requested metrics have been measured.
    \item \textit{Suitable function calls to actions transit requests between states.} These function calls receive fragments of requests, such as architectural specifications of blockchain channels, as inputs. The suitability of these calls depends on whether the functions can process the inputs and whether a request is at a state where such calls are required. For instance, if a requested blockchain network has not been deployed, it is too early to invoke evaluation actions. 
    \item \textit{The execution plan of a request is, therefore, a sequence of action calls that leads to a goal state.}
    \item \textit{A request's state space is implicit} because it depends on NVAL's action pool when it starts to plan for a request. If the state-space of a request is exhausted before reaching a goal state, then the request is unsatisfiable with the current action pool. 
\end{itemize}

We propose to employ a state-space search approach \cite{Russell2010} to solve the action planning problem of NVAL. A state-space search is an effective approach because an agent (NVAL) has (1) full observability over the states, (2) has a set of actions with deterministic effects, and (3) can determine whether a state satisfies the goal. The main idea of state-space search is to systematically traverse the state space to find a goal state with a search algorithm such as breadth-first search. The path taken by the search algorithms becomes the execution plan. Section \ref{sec:planning} presents the problem formulation and solution in details. 

\section{Blockchain Network Design Meta-model}
\label{sec:BND}

\renewcommand{\arraystretch}{1.5}
\begin{figure*}[ht]
    \begin{subtable}{1\textwidth}
        \footnotesize
        \resizebox{\textwidth}{!}{
            \begin{tabular}{@{}llp{0.65\textwidth}@{}}
                \toprule
                \textbf{Vertex Type}    & \textbf{Properties} & \textbf{Description}                                                                                          \\ \midrule
                Blockchain Node (BNode) & Node ID, Node type (Full, Mining, Lightweight, Remote) & A deployed and operational node in a blockchain network                                                       \\
                Blockchain Channel      & Channel ID, Consensus Protocol & A logical channel of a blockchain network consisting of multiple blockchain nodes maintaining the same ledger \\
                Process                 & Description & A business process or technical process that drives the deployment of a blockchain network                    \\
                Participant             & Description, Multiplicity (Individual or multiple users) & A process participant, which can be an individual or an organisation                                          \\
                Computing Node (Node)   & Host name, hardware architecture, OS & A computer that runs a blockchain node                                                                        \\
                Computer Network (Net)  & Net ID, network type, bandwidth & A wired or wireless network that connects computing nodes                                                     \\ \bottomrule
            \end{tabular}
        }
        \caption{Vertices} \label{tbl:BND_metamodel_vertices}
    \end{subtable} \par\bigskip
    \begin{subtable}{1\textwidth}
        \resizebox{\textwidth}{!}{
            \begin{tabular}{@{}lllll@{}}
                \toprule
                \textbf{Edge Type}     & \textbf{From Vertices} & \textbf{To Vertices} & \textbf{Properties}                             & \textbf{Description}                                                                 \\ \midrule
                Participate-in-Process & Participant            & Process              & Role, Authorisation Level                       & Representing a participant's involvement, role, and authorisation level in a process \\
                Control-BNode          & Participant            & BNode                &                                                 & Representing the ownership and control of a participant over a blockchain node       \\
                In-Channel             & BNode                  & Channel              &                                                 & Representing that a blockchain node works in a logical channel                       \\
                Deploy-on-Node         & BNode                  & Node                 &                                                 & Representing that a blockchain node is deployed on a computing node                  \\
                In-Network             & Node                   & Network              & Network interface metadata (e.g., host address) & Representing that a computer is attached to a wired or wireless network              \\ \bottomrule
                \end{tabular}
        }
        \caption{Edges} \label{tbl:BND_metamodel_edges}
    \end{subtable} \par\bigskip
    \begin{subtable}{1\textwidth}
        \footnotesize
        \resizebox{\textwidth}{!}{
            \begin{tabular}{@{}lllp{0.6\textwidth}@{}}
                \toprule
                \textbf{Architectural Structure} & \textbf{Included Vertices}       & \textbf{Included Edges}                            & \textbf{Description}                                                                                                                                       \\ \midrule
                Organisational          & Participants, Processes & Participate-in-Process                    & Describing relationships and power dynamics between participants, which act as constraints of a blockchain network architecture                   \\
                Physical                & Nodes, Networks         & In-Network                                & Describing a targeted hardware infrastructure for deploying a blockchain network, which presents constraints of a blockchain network architecture \\
                Logical                 & BNodes, Channels        & In-Channel, Control-BNode, Deploy-on-Node & Describing a blockchain network's topology and providing the necessary information to deploy it                                                   \\ \bottomrule
            \end{tabular}
        }
        \caption{Architectural structures} \label{tbl:BND_metamodel_structures}
    \end{subtable}
    \caption{BND meta-model for modelling blockchain network architecture} \label{tbl:BND_metamodel}
\end{figure*}

The Blockchain Network Design (BND) meta-model \textit{describes blockchain network architectures as directed property graphs.} This approach is motivated by the following observations about the design decisions underlying blockchain network architectures. Observations (1) to (3) show that blockchain network architecture can be modelled as a directed graph whose nodes and edges carry additional properties (a property graph). Observation (4) shows that a blockchain network architecture must also contain the organisational and physical structures around a blockchain network to provide a context for design decisions. 

\begin{enumerate}
 \item \textit{Some design decisions involve adding new connections between entities.} For instance, the assignment of blockchain nodes to participants can be conceptualised as connections or links between them, specifying ``assigned-to'' relationships.
 \item \textit{Some design decisions add both connections and entities.} For instance, describing a blockchain network topology requires introducing entities such as blockchain nodes and channels. Describing a topology also requires introducing connections between nodes and channels, specifying ``participate-in'' relationships.
 \item \textit{Some design decisions also involve assigning attributes to existing entities and connections.} For instance, the type and configuration of consensus protocols can be captured as attributes of a ``blockchain channel'' entity. Similarly, node type (full, mining, lightweight, remote, interface) can be modelled as an attribute of a ``blockchain node'' entity.
 \item \textit{Some design decisions require information beyond the control of an architect.} For instance, the decision to deploy a blockchain node on a server requires information about the server, such as its specification and network address. Similarly, the decision to assign a blockchain node to a participant requires details about the participant and possibly its involved business processes. The architect controls neither the server nor the participant yet requires to add details to the architecture to provide context for design decisions. 
\end{enumerate}

\textbf{Meta-model:} The BND meta-model defines six vertex types and five edge types that architects can use to build property graphs that represent concrete blockchain network architectures. Figure \ref{tbl:BND_metamodel} presents the types of vertices and edges making up the BND meta-model. The vertices describe the necessary entities for specifying design decisions, including blockchain nodes (BNode), blockchain channels, computing node (Node), computer network (Net), participants, and processes. The edges describe the relationships between entities. For example, an \texttt{in-channel} edge connecting a blockchain node and a blockchain channel shows that the node participates in the channel. Vertices and edges can carry additional properties such as computer hardware architecture, network bandwidth, blockchain node types, and consensus protocol configurations. Architects leverage vertices, edges, and properties to capture all the design decisions defined in the blockchain network design space \cite{Tran2021} and develop blockchain network architectures. In this paper, we denote the resulting property graphs representing blockchain network architectures as \textit{BND models}. They can be serialised as JSON or XML documents for consumption by a software system like NVAL. 

The resulting BND models describe three perspectives of blockchain network architectures simultaneously: organisational, physical, and logical (Table \ref{tbl:BND_metamodel_structures}). \textit{The organisational structure} of a blockchain network captures its \textit{organisational context and requirements}, describing how participants collaborate and their authority level. \textit{The physical structure} describes the underlying computing infrastructure available for running a blockchain network, capturing a network's \textit{physical contraints}. The organisational and physical structures shape the design decisions captured in \textit{the logical structure}, which describes a blockchain network's topology, the deployment of its nodes, and the assignment of its nodes to participants. 

\begin{figure}[ht]
 \centering
 \includegraphics[width=\columnwidth,,trim={0cm 0cm 0cm 0cm},clip]{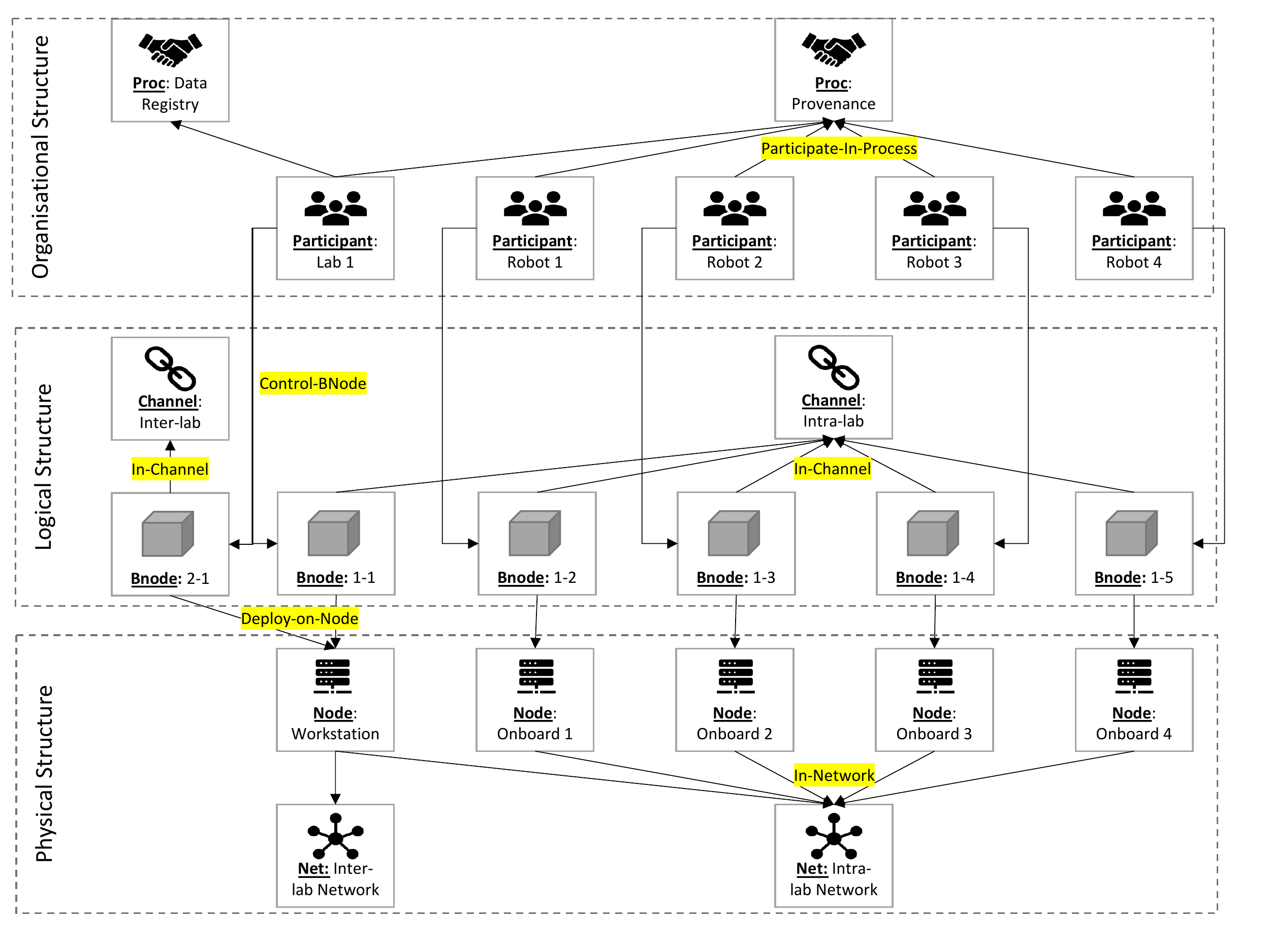}
 \caption{An exemplary BND model of a two-channel blockchain network running on a cluster of five computing nodes.}
 \label{fig:sample_BND}
\end{figure}

\textbf{Example:} Let us demonstrate the meta-model with an exemplary BND model of a two-channel blockchain network in the context of the motivating example (Figure \ref{fig:sample_BND}). The blockchain network serves five participants within the same lab, including a researcher (denoted as \texttt{Lab 1}) and four robots (\texttt{Robot 1} to \texttt{Robot 4}). These participants collaborate on a process (\texttt{Provenance}) that secures and updates the historical records (provenance) of mission objectives and corresponding data. The \texttt{Lab 1} participant also participate in another process called \texttt{Data Registry} with other labs (not shown in the figure for brevity). The details regarding participants and processes are captured in the organisational structure of a BND model, forming the organisational context and requirements. 

The blockchain network in Figure \ref{fig:sample_BND} operates on a hardware infrastructure consisting of a workstation (\texttt{Workstation}) and four onboard computers of the robots (\texttt{Onboard 1} to \texttt{Onboard 4}). The workstation and onboard computers connect via an \texttt{Intra-lab Network} such as a local Wifi network. The workstation also connects to other labs' workstations via an \texttt{Inter-lab Network}. The physical structure of a BND model captures details regarding the hardware, forming the physical context and requirements. 

The blockchain network described in Figure \ref{fig:sample_BND} contains two channels. The first channel, \texttt{Intra-lab}, serves the \texttt{Provenance} process. The channel contains five blockchain nodes (\texttt{1-1} to \texttt{1-5}) that are assigned to five participants and deployed on five computers. The second channel, \texttt{Inter-lab}, serves the \texttt{Data Registry} process. Its blockchain nodes are distributed across all participating labs. One of its nodes, \texttt{2-1}, is assigned to \texttt{Lab 1} participant and deployed on the \texttt{Workstation}. Details regarding the type of blockchain nodes and the configuration of the blockchain channels can be further specified in the properties of relevant entities, which we omit from the figure for brevity and clarity. The logical structure of a BND model captures these design decisions.

\textbf{Modelling Architecture with BND:} Before designing a blockchain network, architects must gather requirements regarding the organisational structure that a blockchain network serves. They also need the details regarding the underlying hardware infrastructure. Based on the gathered information, architects fill up a BND model's organisational and physical structures. Afterwards, architects can capture their design decisions in the logical structure as follows: 
\begin{itemize}
 \item \textit{Topological design decisions} describe how a blockchain network is segmented into channels and the number of blockchain nodes per channel. Operators specify these decisions by (1) adding \texttt{blockchain nodes}, (2) adding \texttt{channels}, and (3) adding \texttt{In-Channel} edges between nodes and their channels. 
 \item \textit{Organisational design decisions} capture the assignment of blockchain nodes to participants. The type of the assigned blockchain node reflects a participant's authorisation level. Operators specify these decisions by (1) adding \texttt{Control-BNode} edges between participants and blockchain nodes and (2) specifying the type of blockchain nodes by setting the \texttt{NodeType} property of blockchain nodes. 
 \item \textit{Consensus design decisions} capture the type and configuration of the consensus protocol used by blockchain channels in the network. Operators specify these decisions by setting the \texttt{ConsensusProtocol} property of channels. 
 \item \textit{Deployment design decisions} capture how blockchain nodes are deployed on targeted hardware infrastructure. Operators specify these decisions by adding \texttt{Deploy-on-Node} edges between blockchain nodes and computing nodes. 
\end{itemize}

\textbf{Parsing Architecture with BND:} BND enables NVAL to process multi-channel blockchain networks in a channel-by-channel fashion. Let us demonstrate how NVAL parses the exemplary BND model presented in Figure \ref{fig:sample_BND} for deployment. The logical structure of the BND model reveals to NVAL that it needs to deploy two blockchain channels (\texttt{Inter-lab} and \texttt{Intra-lab}). Let us take a closer look at the \texttt{Intra-lab} channel. By following its \texttt{In-Channel} edges, NVAL can identify the blockchain nodes that it needs to establish (\texttt{BNode 1-1} to \texttt{BNode 1-5}). By following the \texttt{Deploy-on-Node} edges of the blockchain nodes, NVAL learns where to deploy the blockchain nodes (\texttt{Workstation} and \texttt{Onboard 1} to \texttt{Onboard 4}). By inspecting the properties of the \texttt{Intra-lab} channel, NVAL learns the type of required blockchain platform and configurations. By inspecting the properties of the computing nodes, NVAL learns that it needs to deploy blockchain clients on a powerful workstation running an X64-based processor and multiple resource-constrained onboard computers running ARM-v7 processors. The information regarding the required blockchain platform and the hardware architecture form the requirements used by NVAL to look for suitable automation programs. For instance, assuming that the \texttt{Intra-lab} channel utilises the Proof-of-Authority (PoA) variant of Ethereum, NVAL must look for automation programs that can deploy and configure PoA Ethereum clients on both X64-based and ARMv7-based computers. A similar parsing and processing can be performed for the other channel. 

The ability to parse and comprehend blockchain network architectures also facilitates new architecture validation capabilities. For instance, verifiers can be developed to leverage heuristic or formal analyses to detect misconfigurations, such as deploying a resource-intensive mining node on a resource-constrained onboard computer. 

\section{Composable Actions}
\label{sec:Actions}

NVAL deploys and evaluates blockchain networks by reusing and combining existing automation programs, denoted as \textit{composable actions}, according to the given blockchain network architectures. NVAL acquires composable actions via inner-source (codes from previous projects) or open-source channels (codes from the community). This section presents our proposed scope and architecture of composable actions to facilitate their development and reuse.

\renewcommand{\arraystretch}{1.2}
\begin{table*}[ht]\footnotesize
    \caption{Interface and description features of NVAL's actions} \label{tbl:action_types}
    \resizebox{\textwidth}{!}{
    \begin{tabular}{@{}p{0.1\textwidth}p{0.35\textwidth}ll@{}}
    \toprule
    Action Type & Inputs & Outputs & Description Features \\ \midrule
    Deployer & BND fragments that describe a blockchain channel & An operational blockchain network channel & \begin{tabular}[c]{@{}l@{}}Supporting blockchain platform\\ Supporting types of hardware architecture and OS\\ Supporting mixed hardware architecture\end{tabular} \\ \midrule
    Evaluator & \begin{tabular}[c]{@{}l@{}}A list of required metrics\\ BND fragments of the required channels\end{tabular} & A dataset containing evaluation results & \begin{tabular}[c]{@{}l@{}}Supporting metrics\\ Supporting blockchain platform\end{tabular} \\ \midrule
    Verifier & A request (BND instance and a list of requested metrics) & \begin{tabular}[c]{@{}l@{}}One of the following:\\ - No problem\\ - Recommendations\\ - Warnings\\ - Errors\end{tabular} & Description of the verification written in natural language \\ \bottomrule
    \end{tabular}
    }
\end{table*}

\textbf{Action Scope:} Balancing reusability and abstraction was our goal when defining the scope of composable actions. Based on our experience in developing and refactoring automation scripts for blockchain networks, we proposed to scope NVAL's actions into three types corresponding to three types of activities conducted by NVAL. Table \ref{tbl:action_types} presents the interface and description features of three action types that NVAL leverages to deploy and evaluate blockchain networks. NVAL requires the automation programs contributed by operators via inner-source or open-source channels to conform to these interfaces for interoperability. 
\begin{itemize}
    \item \textit{Deployers} encapsulate the automation logic to \textit{deploy and configure one blockchain channel.} They receive a fragment of a BND model describing a channel as input. NVAL distinguishes and chooses deployers based on the type of blockchain platforms they support, the types of hardware architecture and operating systems they can work with, and whether they can work with multiple types of hardware architecture within one channel (e.g., X64-based and ARMv7-based hosts).
    \item \textit{Evaluators} encapsulate the automation logic to \textit{measure specific evaluation metrics from one blockchain channel.} They receive a list of required metrics and a BND fragment of the targeted blockchain channel as inputs. NVAL distinguishes and chooses evaluators based on the metrics they can measure and the types of blockchain platforms they can work with. 
    \item \textit{Verifiers} encapsulate reusable verifications of incoming requests, such as schema validation and consistency checking. Verifiers are optional actions. We introduced them to capture the potential analyses provided by BND models. 
\end{itemize}

\textbf{Action Architecture:} How to ensure NVAL can understand and execute actions, regardless of their underlying technology, whilst minimising the complexity and implementation effort facing action contributors? We address these challenges by proposing a kernel-based microservice architecture for NVAL's actions. From NVAL's perspective, every action is a self-contained black box accessibly only via a predefined service interface that conforms to ones specified in Table \ref{tbl:action_types}, ensuring that NVAL can understand and invoke actions. Every action is packaged and delivered as a software container that contains all necessary libraries and tools, such as Ansible or Hyperledger Caliper, allowing NVAL to host and operate actions without managing their dependencies. 

\begin{figure*}[th]
    \centering
    \includegraphics[width=0.7\textwidth,,trim={0cm 11cm 7.5cm 0cm},clip]{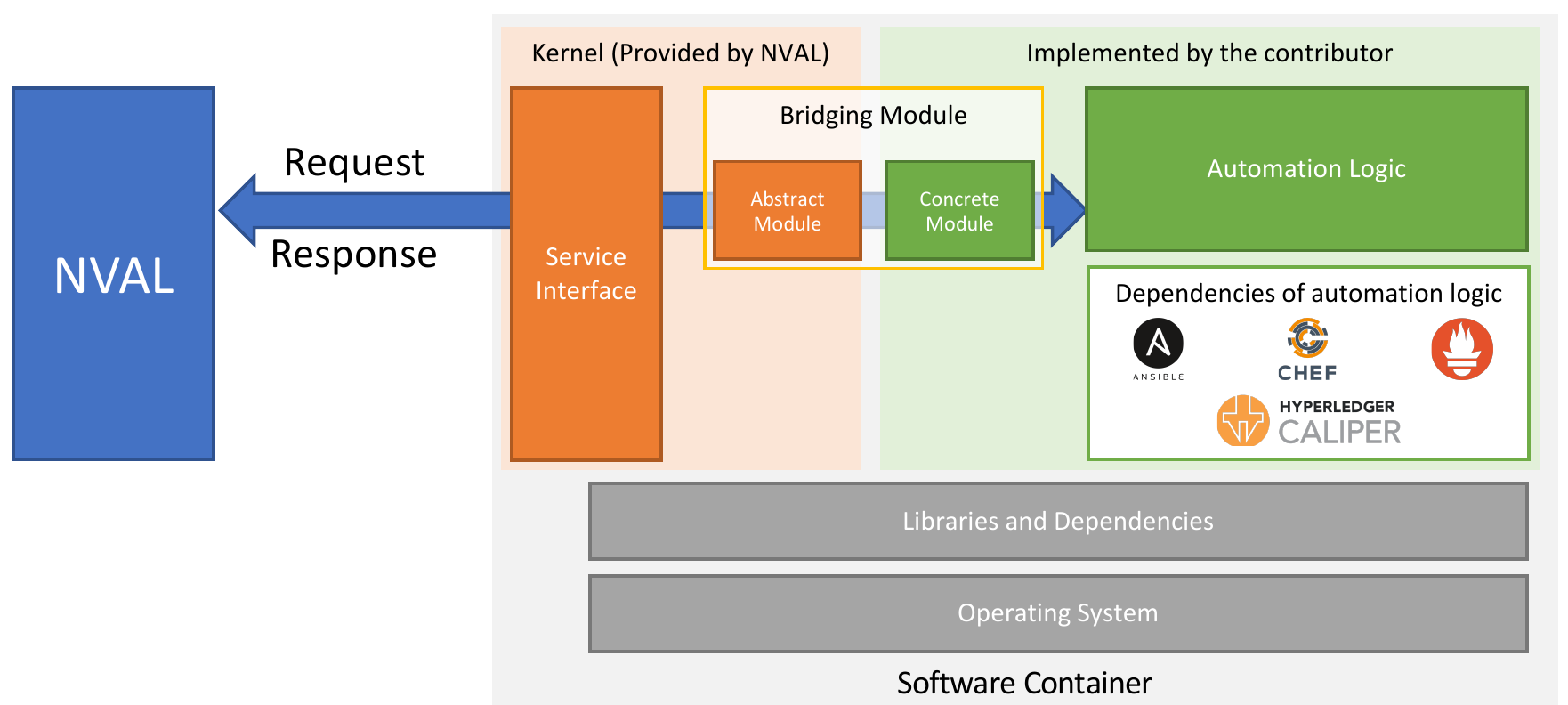}
    \caption{Architecture of a composable action for NVAL}
    \label{fig:NVAL-Action-Architecture}
\end{figure*}

\begin{figure*}[th]
    \centering
    \includegraphics[width=0.7\textwidth,,trim={0cm 0cm 0cm 0cm},clip]{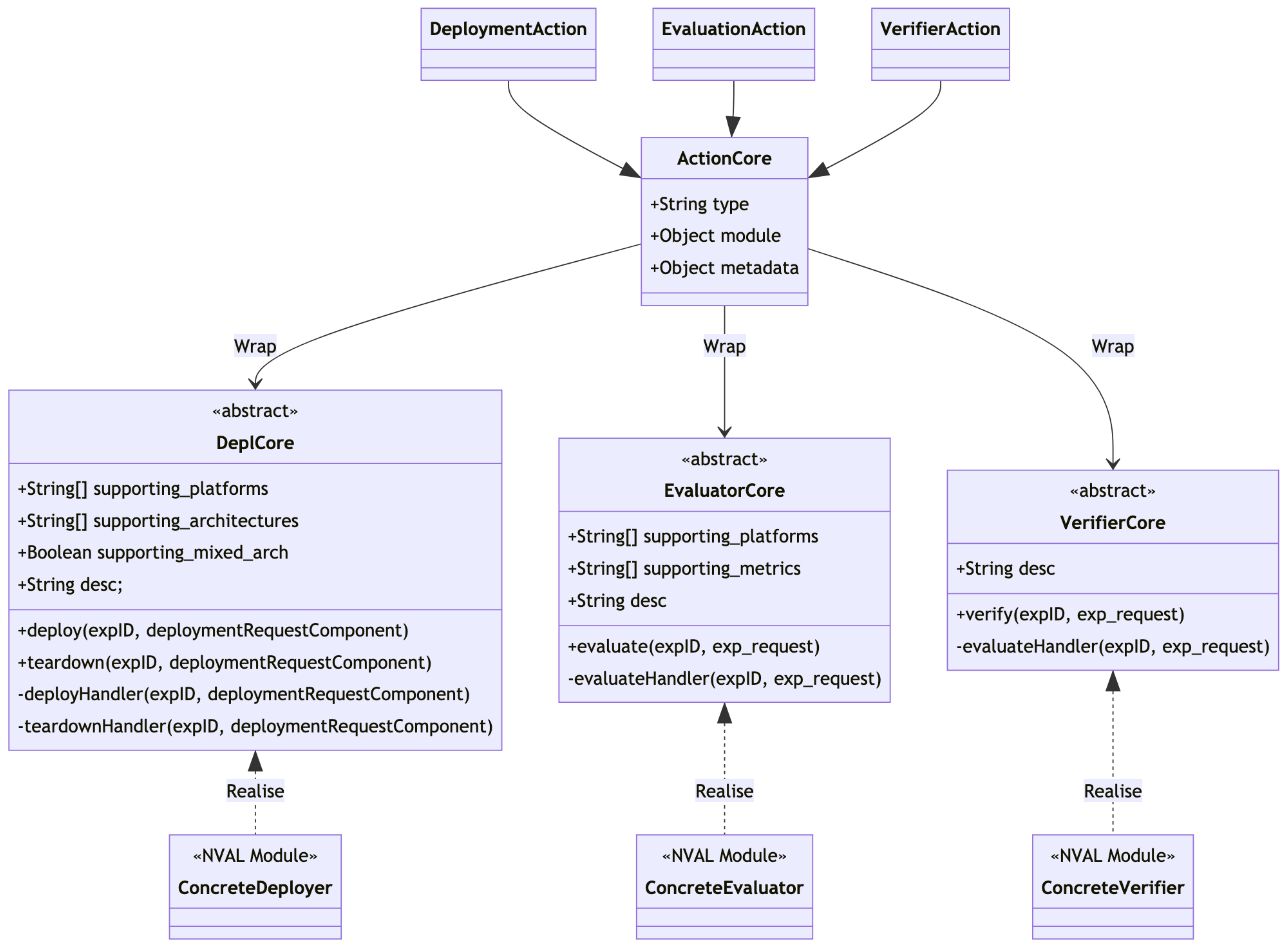}
    \caption{Abstract classes for implementing a bridging module in a NVAL action}
    \label{fig:NVAL-Actions}
\end{figure*}

Developing actions as containerised services undoubtedly introduces additional complexity beyond the scope and focus of contributors. We leverage a kernel-based architecture to abstract the complexity and ensure the compliance of the service implementation. Specifically, we divide the codebase of composable actions into kernel space and contributor space (Figure \ref{fig:NVAL-Actions}). The contributor space contains automation scripts (e.g., Ansible playbooks) and the necessary tools to execute the scripts (e.g., Ansible). The kernel space handles service interface implementation and operation. A bridging module connects kernel and contributor spaces, translating NVAL's service requests to automation script calls. NVAL provides abstract bridging modules (Figure \ref{fig:NVAL-Actions}) that contributors can extend to develop concrete bridging modules. 

From a contributor's perspective, the development process of action is as follows:
\begin{enumerate}
    \item Determining the type of action to develop based on Table \ref{tbl:action_types}
    \item Refactoring or developing automation programs according to the defined scope
    \item Acquiring the kernel from NVAL
    \item Developing a bridging module by subclassing a suitable abstract module (\texttt{DeplCore}, \texttt{EvaluatorCore}, \texttt{VerifierCore}) and implement its abstract functions (e.g., \texttt{deployHandler()}, \texttt{teardownHandler()}). At runtime, NVAL calls these functions though the service interface to reach the automation programs. By implementing these functions, contributors control how NVAL uses their automation programs.
    \item Filling in the action metadata in the bridging module
    \item Packaging the service interface, the bridging module, the automation programs, and necessary utilities into a container. The kernel can provide a template for conducting this step. 
\end{enumerate}

\textbf{Rationale and Counter-Example:} The action scopes are informed by our experiences in developing automation scripts for blockchain networks. Let us consider the following a \textit{counter-example automation script} we developed for the exemplary BND model above. The script received a list of computing hosts as input. Each list entry specifies a blockchain node's network address and channel. The counter-example script instructs Ansible, an automation tool, to (1) deploy one \texttt{Inter-lab} channel using the PoW Ethereum platform, (2) deploy multiple \texttt{Intra-lab} channels with the PoA Ethereum platform, and (3) benchmark all channels using Hyperledger Caliper. From NVAL's perspective, the described automation script is an overly large action that simultaneously performs multiple tasks. We made the following observations:
\begin{enumerate}
    \item \textit{Overly large action scopes hamper reuse:} For instance, we cannot reuse the counter-example script to deploy and evaluate PoA Ethereum channels in a different blockchain network directly because its ability to deploy PoA Ethereum channels is tightly coupled with deploying PoW Ethereum channels and running Caliper benchmarks. 
    \item \textit{Overly small scopes hamper abstraction:} For instance, if we narrow down actions to specific technical steps such as building and deploying an Ethereum client, operators would need an in-depth understanding of Ethereum to use these actions. In other words, we lose the abstraction of implementation details, which is NVAL's primary goal. 
    \item \textit{Deploying and evaluating channels are atomic activities.} From an operator's perspective, deployment and evaluation are either successful or failed. In other words, it does not matter to them whether the failure happens at the first compilation step or the last provisioning step. Therefore, the scope of actions must cover an entire deployment or evaluation step to preserve their abstraction.
    \item \textit{Channels divide the work on a multi-channel blockchain network into independent units.} For instance, we can break the deployment and evaluation of a three-channel blockchain network into three parts; each deploys and evaluates one channel. The same procedure can be used for all channels if they employ similar blockchain technology and request the same evaluation metrics.
    \item \textit{Actions can be distinguished by what channels they can act on and how they do so.} For instance, a deployment action might be able to build and deploy Ethereum nodes only on computers with an ARMv7 processor. This information helps NVAL match this action with blockchain network architectures that feature Ethereum protocol and low-power computers running ARMv7 chips. 
\end{enumerate}

The observations (1) to (3) show that \textit{deployment and evaluation of individual channels} are atomic activities for handling heterogeneous multi-channel blockchain networks. Observation (4) shows that we need to reduce action scopes to individual blockchain network channels rather than an entire network to increase their reusability. Finally, observation (5) shows us how to describe actions to enable their reuse. The proposed action scopes in Table \ref{tbl:action_types} were developed based on these five observations.

\textbf{Composable Actions in Action} Let us demonstrate how the counter-example automation script can be refactored into composable actions. Based on the scopes defined in Table \ref{tbl:action_types}, the script can be refactored into three following actions: 
\begin{itemize}
    \item \textit{Deployer 1} specialises in PoW Ethereum. Assuming that it only works with computers running X64-based processors, we can denote this deployer as $Depl_{(PoW_Eth, X64, True)}$, meaning it supports PoW Ethereum, X64-based hosts and can work with infrastructure containing multiple host types.
    \item \textit{Deployer 2} specialises in PoA Ethereum. Assuming that it can deploy on both ARMv7 and X64 hosts, we can denote this deployer as $Depl_{(PoA_Eth, ARMv7 \wedge X64, True)}$. 
    \item \textit{Evaluator 1} specialises in measuring performance metrics such as latency and throughput from Ethereum channels using the Caliper benchmark suite. We denote this evaluator as $Eval_{(Perf, Eth)}$.
\end{itemize}

NVAL can leverage these actions to handle the exemplary BND model presented in Figure \ref{fig:sample_BND}. The \texttt{Inter-lab} channel matches with the Deployer 1 ($Depl_{(PoW_Eth, X64, True)}$) because the channel runs the PoW Ethereum platform and is deployed on a workstation running X64-based processors, thus requiring the capabilities that Deployer 1 provides. The \texttt{Intra-lab} channel matches with Deployer 2 ($Depl_{(PoA_Eth, ARMv7 \wedge X64, True)}$). The evaluator can be applied to both channels. The execution plan for the exemplary BND model is as follows:
\begin{enumerate}
 \item $Depl_{(PoW_Eth, X64, True)}(Inter-lab)$
 \item $Depl_{(PoA_Eth, ARMv7 \wedge X64, True)}(Intra-lab)$
 \item $Eval_{(Perf, Eth)}(Inter-lab)$
 \item $Eval_{(Perf, Eth)}(Intra-lab)$
\end{enumerate}

\section{Action Planning with State Space Search}
\label{sec:planning}

Action planning denotes the process of finding a suitable sequence of action calls to deploy and evaluate a requested blockchain network archiecture. 

\textbf{Problem Formulation:} We regard the action planning for a request as \textit{a search task on its implicit state-space} with the goal state being the successful deployment and evaluation of a requested blockchain network. We formalise the state-space search problem for action planning as follows.

\begin{definition}[Request]
    A request ($\mathcal{R}$) from an operator contains a blockchain network architecture for deployment and a list of metrics to measure from the deployed network. A request consists of multiple $n$ deployment and $m$ evaluation request components that can be tracked and resolved separatedly: $\mathcal{R}= \left(\mathcal{D}, \mathcal{E}\right)$. \textit{A deployment request component} is an individual blockchain channel to deploy. \textit{An evaluation request components} is a list of metrics to measure from a channel. The state of a request is denoted as $\mathcal{S}_{\mathcal{R}}$.
\end{definition}

\begin{definition}[Action]
    An action is a function that (1) receives a request component, (2) carries out a corresponding task, and (3) returns a new request state. NVAL has three action types: verifier (Eq. \ref{eq:Ver}), deployers (Eq. \ref{eq:Depl}), and evaluators (Eq. \ref{eq:Eval}). The set of actions that NVAL can access to plan and resolve a request is denoted as an \textit{action pool} ($\mathcal{A}$) (Eq. \ref{eq:Actions}).
\end{definition}

\begin{align}
    \mathcal{A} &=\left\{Ver, Depl, Eval \right\} \label{eq:Actions} \\
    Ver &= \left\{ver | ver : \mathcal{R} \mapsto \mathbb{B} \right\} \label{eq:Ver} \\
    Depl &= \left\{depl | depl : \mathcal{D} \mapsto \mathcal{S}_{\mathcal{R}} \right\} \label{eq:Depl} \\
    Eval &= \left\{eval | eval : \mathcal{E} \mapsto \mathcal{S}_{\mathcal{R}} \right\} \label{eq:Eval}
\end{align}

\begin{definition}[Action Planning]
    The action planning problem can be formalised as follows: \textit{given a request $\mathcal{R}$, generate an action sequence $\mathrm{A_\mathcal{R}} = ((a, input) | a \in \mathcal{A}, input \in \mathcal{R})$ such that the request's state $\mathcal{S}_{\mathcal{R}}$ becomes a goal state after applying $\mathrm{A_\mathcal{R}}$.}
\end{definition}

\textbf{Approach:} We apply state-space search to solve the action planning problem. The main idea is employing a search algorithm (e.g., depth-first search, breadth-first search) to systematically explore a request's state space to find a goal state. The path taken by the search algorithms to reach a goal state becomes the execution plan. The traversal of a request's state space requires a state space model and a set of state transition functions that we define below. 

\textit{Request's State Model:} Following \cite{Russell2010}, \textit{we model a request's state space as a tree} whose nodes represent the request's states and edges represent action calls $(a, input)$. A request's state $\mathcal{S}_{\mathcal{R}}$ is a tuple with $3+n+m$ elements (Eq. \ref{eq:State}). The first three elements ($v, d, e$) describe a request's overall status, showing whether it has been verified, deployed and evaluated. The remaining $n+m$ elements describe the status of individual request components, including failed (-1), in-progress (0), and successful (1). The initial state of every request is an all-zero tuple. Its \textit{goal state is an all-one tuple.}

\begin{align}
    \mathcal{S}_{\mathcal{R}} &=\left(v, d, e, req\right) \label{eq:State} \\
    req & \in \{-1, 0, 1\}^{n+m} \\
    v & \in \mathbb{B} \\
    d, e & \in \{-1, 0, 1\}
\end{align}

\textit{State Transition Functions:} The following state transition functions control how a search algorithm traverse a request's state space. 

\vspace{1.5mm}
\noindent \texttt{ACTIONS(s)} determines the possible out-going edges for each state. Each out-going edge represents an action call $(a, input)$. The function returns action calls according to the following rules: 
\begin{itemize}
    \item If $v$ is 0, return all verifiers ($Ver$)
    \item If $d$ is 0, return all deployers ($Depl$) with deployment request components as their input. For instance, if a given blockchain network architecture contains two channels ($ch1, ch2$), and NVAL has two deployers (\texttt{PoW(), PoA()}), then \texttt{ACTION(s)} returns four actions: \texttt{PoW(ch1), PoW(ch2), PoA(ch1), PoA(ch2)}. 
    \item If $d$ is 1 and $e$ is 0, return all evaluators ($Eval$) with evaluation request components as their input, similarly to the above example with deployers.
\end{itemize}

\vspace{1.5mm}
\noindent \texttt{RESULT(s, a)} determines the destination of an out-going edge. In other words, it returns a new state that captures the results of applying an action $a$ to an initial state $s$. The results of this function are according to the following rules: 
\begin{itemize}
 \item If $a \in Ver$, the verification state $v$ is set to 1 to indicate that the verification has been \textit{planned}. 
 \item If $a \in Depl$, the new state depends on whether the deployer in $a$ can handle a given request component. This assessment is done by matching the deployer's descriptive features with the channel's characteristics (e.g., matching the type of supported blockchain platforms with the type of requested blockchain platform). If a deployer can handle a given deployment request component, then the request component's status in $\mathcal{S}^p_{\mathcal{R}}$ is set to 1, indicating that deployment has been \textit{planned}. If all deployment request components have been planned, the overall deployment state $d$ is updated to 1.
 \item If $a \in Eval$, the new state depends on whether the evaluator in $a$ can handle a given evaluation request component. If it can, then the status of the relevant request component is set to 1, indicating that evaluation has been \textit{planned}. If all evaluation request components have been planned, the overall evaluation state $e$ would also be updated to 1.
\end{itemize}

\vspace{1.5mm}
\noindent \texttt{STEP-COST(s, a)} returns the cost of applying an action $a$ on a state $s$. This cost can represent an action's resource consumption (CPU time, memory, network bandwidth), time delay, reliability, or financial costs. For simplicity, we set the STEP-COST uniformly to be one, which means the total cost of an execution plan is the number of actions taken. Future works could explore other STEP-COST to improve the efficiency of the state space search.

\vspace{1.5mm}
\noindent \texttt{IS-GOAL(s)} determines whether a given state $s$ is the goal state. It returns true if all values of a state tuple $\mathcal{S}^p_{\mathcal{R}}$ is 1. If the state space is exhausted before \texttt{IS-GOAL(s)} returns true, the request is deemed unsatisfiable. 

\begin{figure*}[ht]
	\centering
	\includegraphics[width=1\textwidth,,trim={0cm 0cm 0cm 0cm},clip]{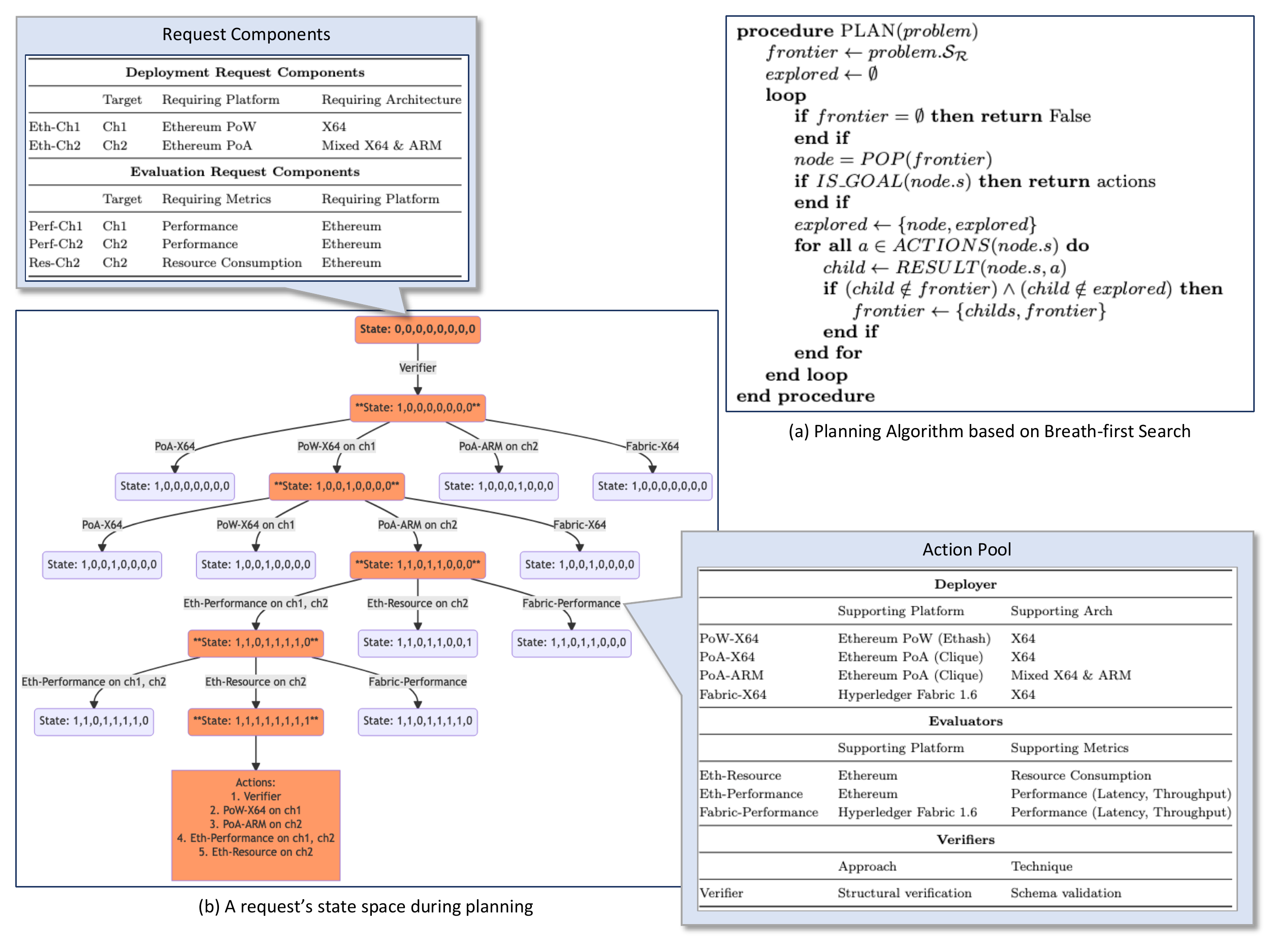}
	\caption{Action planning algorithm and an exemplary planning run}
	\label{fig:NVAL-example}
\end{figure*}

\vspace{1.5mm}
Figure \ref{fig:NVAL-example}a depicts how a breadth-first search algorithm leverages the state transition functions to traverse the state space to search for a goal state. The \texttt{node} object represents a node in the state tree, which contains a state vector (\texttt{s}) and a set of \texttt{actions} leading to it. The \texttt{POP()} removes and returns the last item from an array.

\textbf{State-space Search in Action:} We describe an example that shows how NVAL plans to deploy and evaluate the exemplary BND model described in Section \ref{sec:BND} (Figure \ref{fig:sample_BND}). Recall that the exemplary BND model contains an inter-lab channel and an intra-lab channel, which we denote as \texttt{ch1} and \texttt{ch2}. The first channel (\texttt{ch1}) uses a PoW Ethereum variant to connect workstations of the collaborating labs (X64-based computers). The other channel (\texttt{ch2}) is an intra-lab channel that uses a PoA Ethereum variant to connect robots (ARMv7-based computers) and workstations (X64-based computers) within a lab. Suppose that an operator wants performance metrics (latency and throughput) from \texttt{ch1} and \texttt{ch2} and resource consumption metrics from \texttt{ch2}. In total, the described request can be broken down into two deployment request components and three evaluation request components. Figure \ref{fig:NVAL-example} presents the request components and the action pool available for NVAL to deploy and evaluate the exemplary request.

The state tuple of the sample request consists of eight components (three general states ($v, d, e$), two deployment requests, and three evaluation requests). The request is initially assigned with an all-zero state tuple $(0, 0, 0, 0, 0, 0, 0, 0)$. The breadth-first search algorithm presented in Figure \ref{fig:NVAL-example}a was employed to traverse the state space. Let's consider the expansion of the initial state as an example.
\begin{enumerate}
    \item IS-GOAL(s) is applied on the initial state. It returns false, allowing the search to continue.
    \item ACTIONS(s) is applied on the initial state to identify all available actions. Because $v$ is 0, all verifiers are returned as possible actions.
    \item RESULT(s, a) is applied on the starting state and the verifier action. It returns a new state $(1, 0, 0, 0, 0, 0, 0, 0)$, reflecting that verification has been planned.
    \item STEP-COST(s, a) assigns the cost of the verification to be 1, indicating that one action has been planned. 
\end{enumerate}

The traversal continues until reaching a goal state. All actions leading to that state becomes the experiment plan for the given request. Figure \ref{fig:NVAL-example} depicts the state tree and the search process. 

\section{Evaluation}
\label{sec:Evaluation}

This section presents a case study that aims to evaluate and demonstrate the feasibility of NVAL in a targeted usage scenario. We validated the framework's feasibility in two aspects. Firstly, we demonstrated that NVAL could combine a small set of composable actions to deploy and evaluate a larger set of complex blockchain network architectures. Secondly, we demonstrated that NVAL does not incur significant overheads to the deployment and evaluation process. We formulated these aspects of feasibility into the following research questions:
\begin{itemize}
 \item \textit{RQ1: How effective is NVAL at orchestrating actions for deploying and evaluating diverse blockchain network architectures?}
 \item \textit{RQ2: What is the degree to which NVAL can reuse actions across requests?}
 \item \textit{RQ3: What is the processing time overheads incurred by NVAL?}
\end{itemize}

We chose the case study method to evaluate and demonstrate NVAL's feasibility because it is a powerful and flexible research method to investigate a contemporary phenomenon in its context \cite{Yin2009,Runeson2008}. The studied phenomenon is the feasibility of NVAL. The context in which we study NVAL's feasibility is an architecture development project that employs a proof-of-concept system of NVAL to experiment with various blockchain network architectures. For clarity, we use the term \textit{case} to denote the architecture development project (context) and the term \text{case study} or \textit{study} to denote the evaluation of NVAL via analysing the case. We collected data from the experiment results and NVAL's logs by the end of the case as empirical evidence to answer the research questions. The following sections elaborate on the case, the methodology, and the results.

\subsection{Case Description}

\textbf{Context:} Recall that in the motivating scenario, multiple collaborating research labs wish to leverage a private blockchain to interconnect their robotic testbeds to form a joint experimentation platform. The labs must establish a private blockchain network that bridges various organisations' clouds, workstations, and robots to leverage a private blockchain. Such a network has various potential architectures \cite{Tran2021a}, which offer different trade-offs between quality attributes such as performance (transaction latency and throughput), resource consumption, availability, and security. 

The case of our study is an architecture development project that emerged from this context. It aimed to identify an optimal blockchain network architecture for the joint experimentation platform by empirically evaluating potential architectures using a surrogate testbed infrastructure. The testbed resembles a scaled version of the joint experimentation platform. The project includes \textit{13 experiments}, each of which involves the deployment and benchmarking of a potential blockchain network architecture. The experiments were repeated five times, bringing the total number of deployed blockchain deployed channels to 150, constituting 65 blockchain networks. Most of the deployed blockchain channels produce two evaluation datasets (performance and resource consumption), bringing the total number of datasets to 295. Due to the massive workload, the authors were involved in the project to employ NVAL to support the experiments.

\begin{figure*}[ht]
    \centering
    \includegraphics[width=0.8\textwidth,,trim={0cm 8.5cm 0cm 0cm},clip]{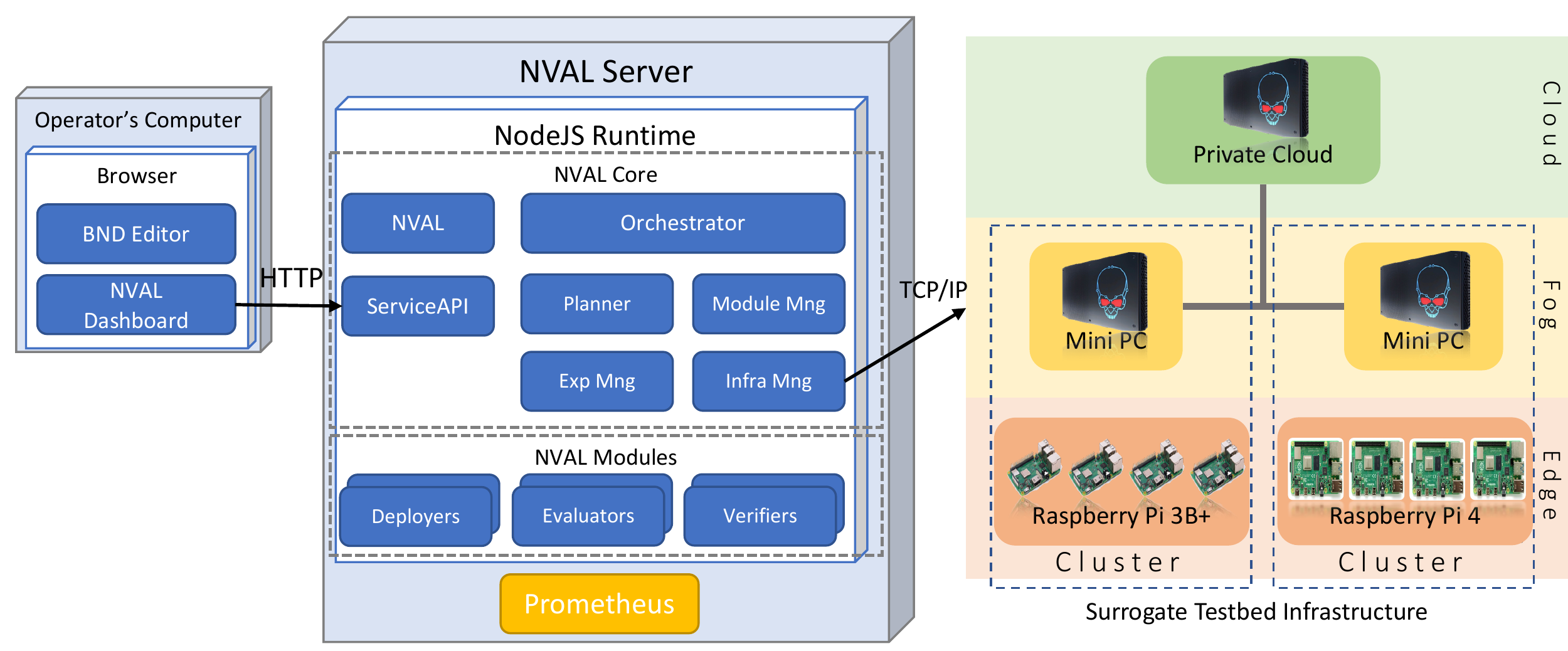}
    \caption{Deployment Structure of NVAL Proof-of-Concept System and the Surrogate Testbed Infrastructure}
    \label{fig:NVAL-Deployment}
\end{figure*}

\textbf{Surrogate Testbed Infrastructure:} The surrogate testbed infrastructure is a scaled version of the joint experimentation platform. The testbed is deployed and operated by the authors to conduct experiments in a controlled environment. Figure \ref{fig:NVAL-Deployment} presents the testbed, which contains two device clusters representing separate robotic testbeds from two research labs. Each cluster contains four \textit{edge devices}, which are resource-constrained computers (Raspberry Pi 3B+ and Raspberry Pi 4) representing robots' onboard controllers. Each cluster contains a powerful workstation acting as a cluster head or base station. The testbed also includes a workstation representing a remote cloud infrastructure that the research labs in the motivating scenario might utilise for data storage and processing. Upon this testbed infrastructure, experiments on blockchain network architectures were conducted.

\textbf{Experimented Architectures:} Twelve potential blockchain network architectures were deployed and evaluated by NVAL. Table \ref{tbl:case_network_arch} presents a detailed breakdown of these architectures in terms of their constituting blockchain channels, where channels' full nodes are deployed (coverage), where channels' mining nodes locate, and the type of blockchain consensus protocol that channels use. The experimented architectures are complex and heterogeneous, featuring both single-channel (D4 to D6) and multi-channel topology (D1 to D3 and D7 to D12) and deploying blockchain nodes on both X64-based computers (base station) and ARMv7-based computers (edge agents).

\begin{table}[ht]
    \centering
    \caption{Blockchain network architectures deployed and evaluated by NVAL in the case study}
    \label{tbl:case_network_arch}
    \resizebox{0.5\textwidth}{!}{
        \begin{tabular}{@{}lllll@{}}
            \toprule
            \textbf{Arch ID} & \textbf{Channel ID} & \textbf{Coverage}               & \textbf{Miner Location}            & \textbf{Consensus} \\ \midrule
            D1              & Ch1, Ch2   & Within edge clusters   & Edge devices              & Ethereum PoA  \\ \midrule
            D2              & Ch1, Ch2   & Within edge clusters   & Fog nodes                 & Ethereum PoA  \\ \midrule
            D3              & Ch1, Ch2   & Within edge clusters   & Fog nodes                 & Ethereum PoW  \\ \midrule
            D4              & Ch1        & Across device clusters & Fog nodes                 & Ethereum PoA  \\ \midrule
            D5              & Ch1        & Across device clusters & Fog nodes                 & Ethereum PoW  \\ \midrule
            D6              & Ch1        & Across device clusters & Fog nodes \& Edge devices & Ethereum PoA  \\ \midrule
            D7              & Ch1, Ch2   & Within edge clusters   & Fog nodes                 & Ethereum PoA  \\
                            & Ch3        & Across device clusters & Fog nodes                 & Ethereum PoA  \\ \midrule
            D8              & Ch1, Ch2   & Within edge clusters   & Fog nodes                 & Ethereum PoW  \\
                            & Ch3        & Across device clusters & Fog nodes                 & Ethereum PoW  \\ \midrule
            D9              & Ch1, Ch2   & Within edge clusters   & Cloud                     & Ethereum PoA  \\
                            & Ch3        & Across device clusters & Cloud                     & Ethereum PoA  \\ \midrule
            D10             & Ch1, Ch2   & Within edge clusters   & Cloud                     & Ethereum PoW  \\
                            & Ch3        & Across device clusters & Cloud                     & Ethereum PoW  \\ \midrule
            D11             & Ch1, Ch2   & Within edge clusters   & Fog node                  & Ethereum PoA  \\
                            & Ch3        & Across device clusters & Fog node                  & Ethereum PoA  \\
                            & Ch4        & Across device clusters & Cloud                     & Ethereum PoA  \\ \midrule
            D12             & Ch1, Ch2   & Within edge clusters   & Fog node                  & Ethereum PoW  \\
                            & Ch3        & Across device clusters & Fog node                  & Ethereum PoW  \\
                            & Ch4        & Across device clusters & Cloud                     & Ethereum PoA  \\ \bottomrule
            \end{tabular}
    }
\end{table}

\textbf{NVAL's Action Pool:} The proof-of-concept system contained four deployers, two evaluators, and one verifier to deploy and evaluate the above blockchain network architectures. Table \ref{tbl:case-action-pool} presents the scopes of these actions. The authors constructed all the composable actions by refactoring existing automation programs previously developed for an IT automation tool called Ansible. \texttt{DeplPoA} was the first developed deployer, resembling the counter-example described in Section \ref{sec:Actions}. The deployer was refined and extended to support more blockchain platforms (Ethereum Proof-of-Work) and hardware architecture (ARMv7), creating \texttt{DeplPoA\_x64\_ARM} and \texttt{DeplPoW\_x64\_ARM}. \texttt{EvalPerformanceEthereum} evaluator is capable of conducting performance benchmarks using the Hyperledger Caliper suite and measuring resource consumption of the underlying hardware using Prometheus. It supports both PoA and PoW variants of Ethereum. The \texttt{Verifier} action is capable of validating incoming deployment requests against a schema to detect syntactical issues.

The action pool of NVAL also contains \texttt{DeplBaseline} and \texttt{EvalBaseline}, which are special actions for determining the baseline resource consumption level of the testbed infrastructure. The baseline deployer deploys a unique blockchain channel called ``baseline'', which does not contain any active blockchain node. The baseline evaluator only works with ``baseline'' channels. It measures the resource consumption metrics of nodes underlying these channels over a long period. 

\begin{table}[ht]\footnotesize
    \centering
    \caption{Action Pool}
    \label{tbl:case-action-pool}
    \resizebox{0.48\textwidth}{!}{
    \begin{tabular}{@{}lll@{}}
    \toprule
    \multicolumn{3}{c}{\textbf{Deployer}} \\ \midrule
    & Supporting Platform & Supporting Arch \\ \midrule
    DeplPoA & Ethereum PoA (Clique) & X64 \\
    DeplPoA\_x64\_ARM & Ethereum PoA (Clique) & Mixed X64 \& ARM \\
    DeplPoW\_x64\_ARM & Ethereum PoW (Ethash) & Mixed X64 \& ARM \\
    DeplBaseline & Baseline & Mixed X64 \& ARM \\ \midrule
    \multicolumn{3}{c}{\textbf{Evaluators}} \\ \midrule
    & Supporting Platform & Supporting Metrics \\ \midrule
    EvalPerformanceEthereum & Ethereum & Resource Consumption \& Performance \\
    EvalBaseline & Baseline & Resource Consumption \\ \midrule
    \multicolumn{3}{c}{\textbf{Verifiers}} \\ \midrule
    & Approach & Technique \\ \midrule
    Verifier & Structural verification & Schema validation \\ \bottomrule
    \end{tabular}
    }
\end{table}

\subsection{Proof-of-Concept Implementation and Deployment}

We developed and deployed a NVAL proof-of-concept system to conduct the case study. Figure \ref{fig:NVAL-Deployment} depicts the deployment structure of a NVAL instance:
\begin{itemize}
    \item \textit{NVAL Server} hosts the framework's software components. It exposes functionality as Web services via a RESTful API. We used the Open API Specification version 3.0 to model the API and employed the Swagger Codegen utility to generate the service interface automatically from the specification. Javascript and NodeJS runtime environment were used to implement NVAL's components.
    \item \textit{Operator's Computer} hosts the client-facing software for creating BND models, submitting requests to NVAL and monitoring the experiments (i.e., deployment and evaluation of a blockchain network). We developed the web application with ReactJS and \texttt{react-digraph}\footnote{https://github.com/uber/react-digraph} libraries.
    \item \textit{Target hardware infrastructures} consist of computers and networks used for deploying and evaluating blockchain networks (Figure \ref{fig:NVAL-Deployment}). NVAL server can reach these computers via SSH. 
\end{itemize}

Both the NVAL server and the client-facing software are deployed on a mobile workstation equipped with a quad-core processor and 16GB of memory. 

\subsection{Evaluation Metrics}

We collected data to answer the research questions from experiment results and log files generated by NVAL's proof-of-concept system. We calculate the following metrics based on the data to quantify the effectiveness and efficiency of NVAL. 

\begin{figure}[ht]
 \centering
 \includegraphics[width=0.30\textwidth,,trim={0cm 0cm 0cm 0cm},clip]{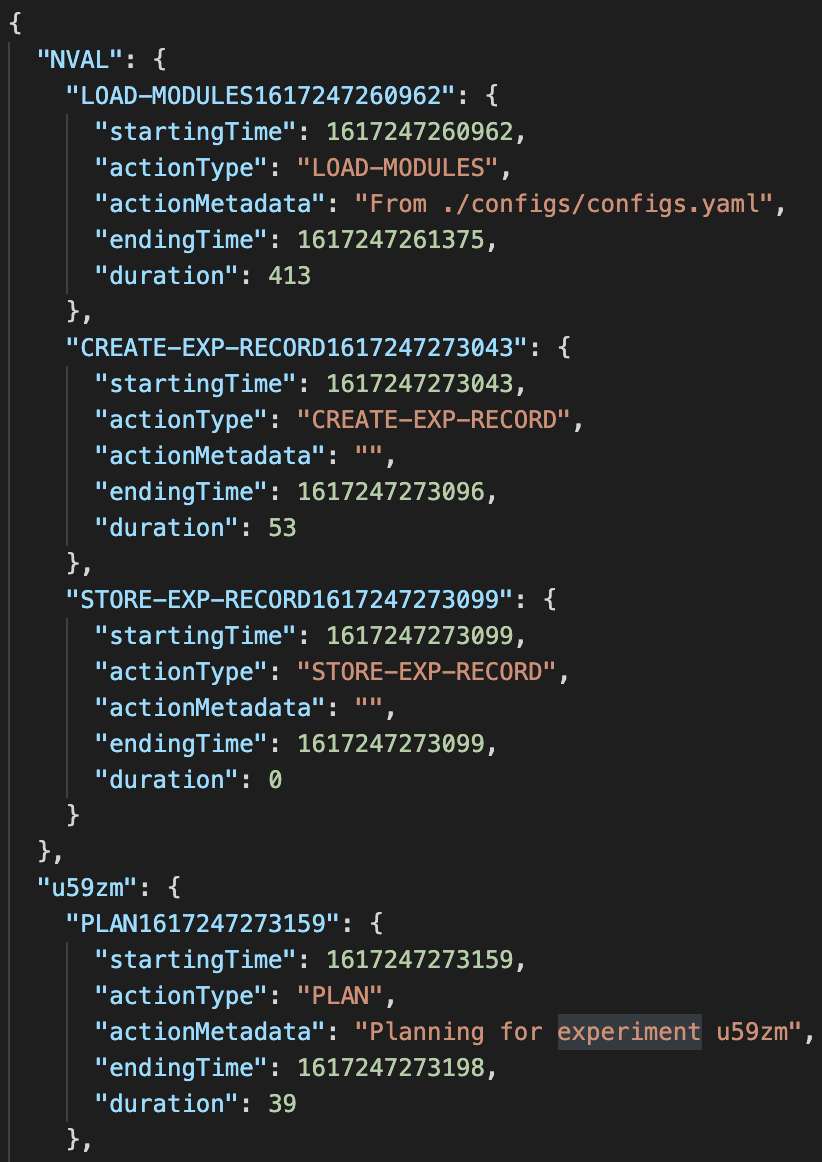}
 \caption{Snippet of a log file generated by NVAL framework}
 \label{fig:sample_log}
\end{figure}

\textbf{Execution Success Rate:} Success rate is the ratio between successful experiment runs and the total number of experiments. An experiment is successful if NVAL applies correct actions in a proper order to deploy a blockchain network correctly and retrieve all the requested performance metrics. The success rate is determined from the experiment results and used to answer RQ1. A high success rate indicates that NVAL is effective at deploying and evaluating blockchain networks. 

\textbf{Per Action Utilisation Rate:} The utilisation rate of an action quantifies the degree to which NVAL can reuse it across multiple requests. This metric is used to answer RQ2. We calculated the utilisation rate from the experiment results in two following ways:
\begin{itemize}
 \item Ratio between the number of experiments in which an action participates and the total number of experiments
 \item Ratio between the number of times NVAL invokes an action, and the total number of action calls NVAL performs across the experiments
\end{itemize}

\textbf{Processing Time Overheads:} This metric quantifies the amount of experiment time spent on NVAL-specific activities such as creating execution plans and invoking actions. We calculated this metric by analysing the NVAL's log files that show us starting and ending times of every activity conducted within NVAL (Figure \ref{fig:sample_log}). This metric is used to answer RQ3. Lower overheads indicate a higher level of efficiency. 

\subsection{Results}

\textbf{RQ1 - The Effectiveness of NVAL:} NVAL framework successfully planned and conducted five rounds of 13 experiments (twelve architecture and one baseline), leading to the deployment and evaluation of 150 blockchain channels, constituting 65 blockchain networks. All evaluation datasets were usable and free of anomalies. We also did not observe any false-negative from the planning mechanism as the action pool was adequate to satisfy all the given requests. Given these observations, we reached the following conclusion for RQ1:

\vspace{2mm}
\noindent\fbox{%
 \parbox{\columnwidth}{%
 RQ1: Given an adequate action pool, NVAL plan and conduct experiments accurately for 100\% of the given experiment requests. 
 }%
}

\vspace{2mm}
\textbf{RQ2 - Utilisation Rate of Composable Actions:} NVAL's action pool served 365 invocations across five rounds of 13 experiments. Table \ref{tbl:utilisation_rate_experiments} presents the utilisation rate of each action across the experiments. Table \ref{tbl:utilisation_rate_action_calls} breaks down the utilisation of each action, showing the number of times it was called across the experiments. 

\begin{figure*}[ht]
 \begin{subtable}[ht]{\textwidth}
 \centering
 \resizebox{0.6\textwidth}{!}{
 \begin{tabular}{@{}lcc@{}}
 \toprule
 & Number of utilising experiments & Utilisation rate \\ \midrule
 DeplBaseline & 1 & 8\% \\
 EvalBaseline & 1 & 8\% \\
 DeplPoA & 3 & 23\% \\
 DeplPoW\_X64\_ARM & 5 & 38\% \\
 DeplPoA\_X64\_ARM & 5 & 38\% \\
 EvalPerformanceEthereum & 12 & 92\% \\
 Verifier & 13 & 100\% \\ \bottomrule
 \end{tabular}
 }
 \caption{Utilisation of actions in terms of experiment participation} \label{tbl:utilisation_rate_experiments}
 \end{subtable} \par\bigskip
 \begin{subtable}[ht]{\textwidth}
 \centering
 \resizebox{\textwidth}{!}{
 \begin{tabular}{@{}lcccccccccccccc@{}}
 \toprule
 \textbf{} & \textbf{Baseline} & \textbf{D4} & \textbf{D6} & \textbf{D5} & \textbf{D3} & \textbf{D2} & \textbf{D1} & \textbf{D10} & \textbf{D7} & \textbf{D9} & \textbf{D8} & \textbf{D11} & \textbf{D12} & \textbf{Grand Total} \\ \midrule
 DeplBaseline & 1 & & & & & & & & & & & & & 1 \\
 EvalBaseline & 1 & & & & & & & & & & & & & 1 \\
 DeplPoA & & & & & & & & & 3 & 3 & & 2 & & 8 \\
 DeplPoA\_X64\_ARM & & 1 & 1 & & & 2 & 2 & & & & & 2 & & 8 \\
 DeplPoW\_X64\_ARM & & & & 1 & 2 & & & 3 & & & 3 & & 4 & 13 \\
 Verifier & 1 & 1 & 1 & 1 & 1 & 1 & 1 & 1 & 1 & 1 & 1 & 1 & 1 & 13 \\
 EvalPerformanceEthereum & & 1 & 1 & 1 & 2 & 2 & 2 & 3 & 3 & 3 & 3 & 4 & 4 & 29 \\ \midrule
 \textbf{Grand Total} & \textbf{3} & \textbf{3} & \textbf{3} & \textbf{3} & \textbf{5} & \textbf{5} & \textbf{5} & \textbf{7} & \textbf{7} & \textbf{7} & \textbf{7} & \textbf{9} & \textbf{9} & \textbf{73} \\ \bottomrule
 \multicolumn{15}{l}{\small The presented invocation counts are from one experiment round. Five rounds were conducted.}
 \end{tabular}
 }
 \caption{Utilisation rate of actions in terms of total invocations} \label{tbl:utilisation_rate_action_calls}
 \end{subtable}
 \caption{Utilisation level of actions in the case}
 \label{tbl:utilisation_rate}
\end{figure*}

The data shows that actions with broad functional scopes can support (i.e., being utilised by) more experiments. For instance, 92\% of the experiments utilised \texttt{EvalPerformanceEthereum} because it can measure both performance and resource consumption, and it supports both protocol variants of Ethereum. Critical actions that are universally applicable, such as \texttt{Verifier} have almost perfect reusability with NVAL. On the other hand, actions with low demand and limited scopes, such as \texttt{DeplBaseline} and \texttt{EvalBaseline}, had a low utilisation rate. 

An interesting observation from the data is that \textit{multiple actions with limited scopes can be combined to cover as many or even more experiments than larger actions.} For instance, the \texttt{DeplPoA} deployer supports only 23\% of the experiments by itself. However, the combination of three deployers working together was able to deploy all 12 blockchain network architectures. This observation suggested the feasibility of NVAL's automation approach of reusing and combining existing automation programs to support future requests. We reached the following conclusion for RQ2: 

\vspace{2mm}
\noindent\fbox{%
 \parbox{\columnwidth}{%
 RQ2: NVAL can reuse an action pool to deploy and evaluate new blockchain networks. It can orchestrate a small set of actions to serve a substantially larger set of requests. 
 }%
}

\begin{figure*}[ht]
    \begin{subtable}[ht]{\textwidth}
    \centering
    \resizebox{0.8\textwidth}{!}{
    \begin{tabular}{@{}lccc@{}}
    \toprule
    & Avg. Duration (milisec.) & Max Duration (milisec.) & Std. Dev (milisec.) \\ \midrule
    \textbf{STORE-EXP-RECORD} & \textbf{0.48} & \textbf{3} & \textbf{0.62} \\
    VERIFY & 2.49 & 18 & 2.94 \\
    \textbf{ORCHESTRATION-OVERHEADS} & \textbf{5.66} & \textbf{28} & \textbf{4.99} \\
    \textbf{CREATE-EXP-RECORD} & \textbf{42} & \textbf{78} & \textbf{9.23} \\
    \textbf{PLANNING-OVERHEADS} & \textbf{47.37} & \textbf{85} & \textbf{19.38} \\
    LOAD-MODULES & 441 & 761 & 80.68 \\
    DEPLOY & 40377 & 164084 & 25187.3 \\
    EVALUATION & 440864 & 791532 & 169850.88 \\ \bottomrule
    \end{tabular}
    }
    \caption{Durations of activities carried out by NVAL}
    \label{tbl:durations}
    \end{subtable} \par\bigskip
    \begin{subtable}[ht]{\textwidth}
    \centering
    \resizebox{0.8\textwidth}{!}{
    \begin{tabular}{@{}lccc@{}}
    \toprule
    & Avg. Duration (milisec.) & Max Duration (milisec.) & Std. Dev. (milisec.) \\ \midrule
    ACTUAL-DEPLOYMENT-EVALUATION & 436675 & 861352 & 168265.33 \\
    TOTAL-OVERHEADS & \textbf{95.5 (0.022\%)} & 176 & 29.9 \\ \bottomrule
    \end{tabular}
    }
    \caption{Total overheads of NVAL against the actual runtime}
    \label{tbl:overheads}
    \end{subtable}
    \caption{Durations of NVAL activities and the framework's overheads.}
    \label{fig:durations_overheads}
\end{figure*}

\vspace{2mm}
\textbf{RQ3 - Processing Time Overheads:} Figure \ref{fig:durations_overheads} presents the durations of various activities conducted by NVAL. The time spent on all activities beside \texttt{VERIFY}, \texttt{DEPLOY}, and \texttt{EVALUATION} are considered overhead. The overhead can be classified into bootstrap overhead and experiment overhead. \textit{Bootstrap overhead} is caused by the \texttt{LOAD-MODULES} activity that is performed once when NVAL boots. We found bootstrap overhead to be around half a second (441 ms). 

\textit{Experiment overhead} is caused by activities related to recording, planning, and orchestrating the deployment and evaluation. This overhead incurs whenever an experiment is conducted. We found the experiment overhead to be around a tenth of a second (95.5 ms), which is minuscule compared to the time required to deploy and benchmark a blockchain network, ranging from 7 to 14 minutes. Based on these findings, we reached the following conclusion for RQ3: 

\vspace{2mm}
\noindent\fbox{%
 \parbox{\columnwidth}{%
 RQ3: NVAL's overheads are negligible compared to the time required to deploy and evaluate blockchain networks.
 }%
}

\section{Discussions}

\subsection{Benefits and Usage Scenario}

The NVAL framework provides practitioners and researchers with the ability to rapidly deploy and evaluate heterogeneous multi-channel blockchain networks. By leveraging BND meta-model, composable actions, and state-space search, the framework can deploy and evaluate blockchain network directly based on their architecture whilst abstracting procedural details and low-level configurations from users. This architecture-driven automation approach benefits practitioners and researchers by allowing them to focus on desiging blockchain network architecture, a valuable and hard-to-automate activity. Automation of implementation and configuration steps also prevent errors such as missing steps and misconfigurations, reducing time and effort in blockchain network deployment and evaluation. Another benefit of NVAL is the capturing and sharing of procedural knowledge regarding deployment and evaluation activities. By capturing this knowledge explicitly as software artefacts (composable actions) rather than documents, practitioners and researchers can version control, test, refine, and employ the knowledge rapidly across multiple organisations. Finally, by enabling practitioners to rapidly deploy and evaluate their blockchain network architectures, NVAL can shorten the feedback loop and enable practitioners to conduct more design iterations, which ultimately lead to better architectures \cite{bass2003software}. 

Unlike existing blockchain automation framework and services, NVAL offers practitioners complete control and transparency over the framework's capability. NVAL is designed such that it can be deployed and operated by individual practitioners on consumer hardware, as demonstrated in the case study. These practitioners maintain a complete control over the capability of their NVAL instances as they can choose the actions making up the action pool of their instances.  The kernel-based approach employed in these composable actions facilitate the development and sharing of new NVAL's capabilities, without relying on permission or approval from any central authority such as service providers or project maintainers. These mechanisms together enable NVAL to keep up with the ever changing landscape of blockchain technology, increasing its transparency and longevity. 

\subsection{NVAL and Beyond}

The case study has shown NVAL's effectiveness in reusing and composing a small pool of automation programs to deploy and evaluate diverse and complex blockchain network architectures. However, our approach and the current implementation still have some drawbacks. This section analyses these drawbacks and presents some potential solutions, constituting future research. 

Firstly, because NVAL reuses existing automation programs to deploy and evaluate blockchain networks, its capability is limited by the breadth of actions it can access. Therefore, the prime challenge of NVAL is ensuring a steady supply of composable actions for both existing and emerging blockchain technologies. The interfaces and the kernel-based architecture proposed in Section \ref{sec:Actions} lay a foundation for practitioners to capture and share their procedural knowledge as composable actions. Future research could focus on the automated transformation of the existing automation scripts and natural language instructions into composable actions. As of February 2022, we found 52 open-source repositories on automated blockchain network deployment with Ansible, 67 repositories with Terraform, and 35 repositories with other scripting languages on GitHub. The official documentation of many blockchain implementations also provide instructions for deploying private networks\footnote{\url{https://geth.ethereum.org/docs/interface/private-network}}. These automation scripts and tutorials represent raw materials for NVAL's actions. 

Secondly, the current version of NVAL does not feature extensive verifications of the contributed actions beyond manual inspections by practitioners. The rationale of this design is the assumption that actions primarily come from trusted inner-sources, and the community verifies open-source actions before they reach an end-user. This model might not scale when NVAL's actions proliferate and increasingly become targets of malicious actors who want to leverage them as an attack vector to infrastructures. Future research could develop a resilient and decentralised registry for discovering and reporting malicious actions. The \texttt{STEP-COST()} function could leverage the ratings from this registry to adjust the planning procedure. 

Thirdly, the current version of NVAL only employs Unicode strings and a substring detection mechanism to describe and match actions with architectural specifications written in BND. For instance, if a blockchain channel is deployed on X64-based computers and uses Ethereum PoW blockchain protocol, NVAL would match the channel with any action containing \texttt{X64} and \texttt{PoW} in its meta-data. The case study demonstrated that the simple approach is effective in matching channels with actions \textit{based on their functionality.} However, substring detection might not be adequate for sophisticated action matching that takes into consideration non-functional attributes of actions such as performance, reliability, and security. Future research could explore sematic descriptions and matching mechanisms. For instance, ontologies could replace the Unicode string to extend the description of actions to cover non-functional requirements. Ontology-based query can replace substring detection in matching channels with actions. 

Finally, NVAL currently does not automatically feed the execution phase's output to the planning phase. Practitioners must inspect the logs manually when a deployment fails and determine whether the failure is due to an exception or an action's failure. If they detect that a composable action is the root cause, they need to remove it from the pool manually. Future research could explore techniques from self-adaptive software systems, such as MAPE-K feedback loops \cite{IBM2006} to enable NVAL to observe and update its action pool automatically. 

\subsection{Threats to Validity}

\subsubsection{Construct validity}

Construct validity concerns whether our case study design successfully assesses NVAL's effectiveness and efficiency in real-world usage. We mitigated this threat by following an established guideline \cite{Runeson2008}, which requires that case selection and data collection follow research objectives.

The selected case supports our research objectives because (1) it is a real-world use case of NVAL, (2) it requires the rapid deployment and evaluation of many blockchain networks, and (3) these blockchain networks have diverse and complex multi-channel architectures. 

The data collection process was also driven by the objectives of assessing NVAL's effectiveness and efficiency. We chose success and failure rates to quantify NVAL's effectiveness because these metrics reflect the framework's ability to correctly use an existing action pool to deploy and evaluate blockchain networks. We chose time overheads to quantify NVAL's efficiency because time is the most tangible and significant resource that NVAL consumes, given that it is not computing-intensive.

\subsubsection{External validity}

External validity concerns whether the effectiveness and efficiency that NVAL shows in the case study can be generalized. Whilst the case study method does not aim to provide statistically significant conclusions \cite{Runeson2008}, we attempted to broaden the generalization of our results by choosing a case that features diverse blockchain network architectures extracted from the literature. The variety of network topology and the inclusion of edge devices and fog nodes expand our generalization scope beyond enterprise blockchain use cases, covering the emerging use cases in cyber-physical systems and edge computing.

\subsubsection{Reliability}

Reliability concerns whether case study results depends on the researchers carrying out the study. We mitigated threats to reliability by the following strategies. Firstly, we avoided subjective data such as NVAL's usability and focused on objective data gathered from the source code and runtime logs. Secondly, we only employed commodity hardware for NVAL and the experiment infrastructure to prevent skewing the efficiency results due to exclusive, high-performing computers. 

\section{Related Work}


This paper addresses the problem of automating blockchain network deployment and evaluation, which contributes to the Blockchain Oriented Software Engineering (BOSE) research \cite{Porru2017,Chakraborty2018,Ortu2019} and aligns with three research areas: blockchain network architecture, blockchain network deployment, and evaluation. 

\begin{table*}[ht]
    \centering
    \caption{Comparison between NVAL and the existing automation frameworks}
    \label{tbl:comparison_related_work}
    \resizebox{\textwidth}{!}{
        \footnotesize
        \begin{tabular}{@{}p{0.17\textwidth}|p{0.22\textwidth}p{0.15\textwidth}p{0.22\textwidth}p{0.1\textwidth}p{0.1\textwidth}p{0.2\textwidth}p{0.1\textwidth}p{0.1\textwidth}@{}}
            \toprule
            Automation Framework            & Description                                                                                                              &  Technology                                 & Input                                                                                             & Support Heterogeneous Blockchain Networks & Support Evaluation & Supported Blockchain Implementations                             & Supported Infrastructures & End-user Extensibility \\ \midrule
            HL Composer                     & Utility for deploying business abstraction such as assets over an existing HL Fabric network. Depricated in 08/2021.     & NodeJS                                                & Business network models (assets, participants, identities)                                            &                                              &                       & HL Fabric                                                        & Cloud                     &                        \\
            PlaTIBART (Walker et al., 2017) & A platform for deploying Ethereum clients wrapped by software agents that handle monitoring and off-chain communication. & Python + SSH                                          & Domain-specific description of Ethereum network configurations and placement of Ethereum nodes        &                                              &                       & Go-Ethereum                                                      & Any                       &                        \\
            Malik et al., 2019              & An automation script for deploying and configuring an HL Sawtooth client on a computer.                                  & Shell script                                          & Practitioners manually place the script on relevant computers and specify node type ot the script.    &                                              &                       & HL Sawtooth                                                      & Any                       &                        \\
            uBaaS (Lu et al., 2019)         & Deployment-as-a-Service for private and consortium blockchain networks                                                   & Infrastructure-as-Code                                & Network-specific configurations (e.g., PoW difficulty level) and IP address for each blockchain node. &                                              &                       & Ethereum, HL Fabric                                              & Cloud                     &                        \\
            NutBaaS (Zheng et al., 2019)    & One-click network deployment based on user-defined profiles                                                              & Unspecified                                           & Unspecified                                                                                           & Yes                                          &                       & Ethereum, HL Fabric, Filecoin                                    & Cloud                     &                        \\
            MixBytes Tank                   & A benchmark suite that can deploy blockchain nodes on private clouds in addition to running benchmarks                   & Python + Ansible                                      &                                                                                                       &                                              & Yes                   & Polkadot, Haya                                                   & Cloud                     &                        \\
            MADT                            & A large-scale IP network simulator that supports deployment of distributed applications (blockchain networks)            & Python                                                & Simulated network configuration                                                                       &                                              & Yes                   & Any                                                              &                           &                        \\
            HL Bevel                        & Utility for deploying containerised blockchain network on any Kubernetes cluster                                         & Ansible + Helm + Kubernetes                           & Mostly low-level and technology-specific configurations (consensus type, port, network address, etc.) &                                              &                       & Corda, HL Fabric, HL Indy, Quorum                                & Cloud                     &                        \\ \midrule
            \textbf{NVAL}                   & \textbf{Architecture-driven, community-supported framework for automating blockchain network deployment and evaluation}    & \textbf{NodeJS + Any technology for automation programs} & \textbf{Platform-agnostic, high-level architecture specifications}         & \textbf{Yes}                                 & \textbf{Yes}          & \textbf{Any (Prototype support Ethereum with Ethash and Clique)} & \textbf{Any}              & \textbf{Yes}           \\ \bottomrule
            \end{tabular}
    }
\end{table*}

\subsection{Blockchain Network Deployment}

Automated blockchain network deployment has been investigated by both academia \cite{Walker2017,Malik2019,Lu2019,Zheng2019,Gorski2021} and industry \cite{Hyperledger2019,Hyperledger2021,MixBytes2021,MADT2020} to abstract the time-consuming \cite{Malik2019}, and error-prone deployment process \cite{Lu2019}, allowing practitioners to focus on application development instead of learning networks and protocols \cite{Zheng2019}. Automation has also been sought to make blockchain network deployment repeatable to support regression tests \cite{Walker2017} and continuous delivery \cite{Gorski2021}.

Hyperledger Composer \cite{Hyperledger2019} is amongst the first tools that recognise and address the complexity of establishing a private blockchain. It helps practitioners implement organisational design decisions by providing them with models and tools to deploy business abstractions such as assets and participants on an existing Hyperledger Fabric network. The Composer was deprecated in 2021.

The automation tools released after Composer have shifted their focus to blockchain networks' logical and physical design decisions. Many of these tools feature some form of domain-specific languages to capture the topology and configurations of the requested blockchain networks. The PlaTIBART platform \cite{Walker2017} is among the first to tackle \textit{blockchain network} deployment, focusing on Ethereum. It proposes a domain-specific language to describe Ethereum's configurations, such as network ID and mining difficulty level. A Python-based network manager uses the configuration files to deploy Ethereum networks on connected computers. Hyperledger Bevel \cite{Hyperledger2021}, formerly known as the Blockchain Automation Framework, is a state-of-the-art tool for rapidly deploying production-ready blockchain networks, supporting Fabric, Corda, Indy, and Quorum blockchain protocols. It uses Ansible, Helm, and Kubernetes to rapidly provision containerised blockchain networks on any Kubernetes cluster. Bevel's inputs are low-level and technology-specific configurations such as consensus types, network addresses and ports. Gorski et al. \cite{Gorski2021} proposed an approach similar to Bevel, but relying on UML deployment models rather than domain-specific models as inputs. Malik et al. \cite{Malik2019} bypassed blockchain network configuration files by proposing a shell script that practitioners manually place on targeted computers and invoke to deploy blockchain nodes. Alternatively, Li et al. \cite{Li2022} proposed an algorithm automate the design and deployment of blockchain network based on the characteristics of sensor data and the underlying edge computing network. 

Many Blockchain-as-a-Service (BaaS) platforms such as uBaaS \cite{Lu2019}, NutBaaS \cite{Zheng2019}, RBaaS \cite{Cai2022}, AWS Managed Blockchain, Azure BaaS, and IBM Blockchain Platform also provide deployment automation as a prominent feature. They generally provide a ``one-click deployment'' utility for spawning a blockchain network across cloud-based virtual machines. This deployment utility is generally vendor-locked in terms of the supported blockchain technologies and computing nodes \cite{Lu2019,Cai2022}. Few offer any form of extensibility besides promised first-party extensions. A recent study on DevOps practices for blockchain-based software engineering found that while BaaS usage is becoming more common, vendor-specific limitations in terms of the supported blockchain platforms and hardware limit their adoption in blockchain-based software projects \cite{Woehrer2021}.

A few blockchain benchmark tools also support blockchain network deployment. For instance, MixBytes Tank \cite{MixBytes2021} helps practitioners deploy a private Polkadot or Haya blockchain on private clouds before running a benchmark. MADT \cite{MADT2020} is a Python-based network simulator that allows practitioners to deploy a private blockchain on the simulated network for development and evaluation. 

Table \ref{tbl:comparison_related_work} compares NVAL with the existing work. NVAL shares the core approach with PlaTIBART and Bevel, which involves capturing blockchain network architectures as digital artefacts and employing a centralised network manager to realise the requested blockchain networks. Unlike the existing system, NVAL's meta-model (Section \ref{sec:BND}) enables architects to specify blockchain network architectures at a design-decision-level rather than a platform-configuration-level and allows them to capture multi-channel heterogeneous blockchain network architecture. The flexibility of blockchain network architecture input is facilitated by community-contributed automation programs (Section \ref{sec:Actions}) and an automated planning approach (Section \ref{sec:planning}). As demonstrated in the case study, NVAL can combine multiple automation programs to process complex multi-channel heterogeneous blockchain networks at runtime. Moreover, end-users such as blockchain network operators can control and extend the framework's capability by acquiring and adding actions to their NVAL instance rather than relying on a project owner or service provider. Besides deployment, NVAL also supports evaluation automation by introducing the \texttt{evaluator} action type to capture evaluation automation tasks and embedding evaluation requests and states into NVAL's planning procedure. Finally, the case study demonstrates that NVAL only requires SSH connections to target computers to deploy and evaluate blockchain networks. Therefore, it can support any infrastructure rather than be constrained to a specific cloud. 

\subsection{Blockchain Network Architecture}

This paper also contributes to the architectural research of blockchain applications and networks. This line of research started from the recognition of blockchain as a software connector type that bridges software components across organisations in a decentralised and trustless manner \cite{xu2016blockchain,Xu2019}. Recent research on the software architecture aspects of blockchains has focused on architectural tactics \cite{Wessling2018} and patterns \cite{Xu2018} to design on-blockchain smart contracts and the software systems surrounding them. 

Whilst blockchain networks are also software systems, research on blockchain network architectures have received less focus. Xu et al. \cite{Xu2017} proposed a taxonomy of blockchain networks to support practitioners in choosing a blockchain for their use cases. Tran et al. \cite{Tran2021a} was amongst the first to investigate the possible design decisions in establishing a blockchain network. Recent research has identified design patterns of blockchain networks for edge computing use cases, highlighting the prevalence of multi-channel heterogeneous blockchain networks \cite{Tran2021a}. 

This paper advances the research on blockchain network architecture by proposing a meta-model for capturing blockchain network architectures as computer-understandable artefacts. This meta-model is based on the design decisions identified by Tran et al. \cite{Tran2021a}. Future research, such as formal verification and recommendation of blockchain network architecture, can leverage the proposed meta-model. 

\subsection{Blockchain Network Evaluation}

Blockchain network evaluation can be classified into benchmark and simulation. Benchmarks assess a blockchain network's quality after deployment by subjecting the network to standardised workloads and measuring quality attributes of interest. Benchmarks can be conducted manually as demonstrated on the Quorum \cite{Baliga2018} and Red Belly \cite{Crain2018}, and Fabric \cite{Wickboldt2019} blockchain platforms. As proprietary benchmarks can be difficult to compare, researchers and industry have developed blockchain benchmark suites. BLOCK BENCH \cite{Dinh2017} is among the earliest benchmark automation suites. It provides workloads for benchmarking both end-to-end performance and individual layers such as consensus and processing. Pongnumkui et al. \cite{Pongnumkul2017} extended the idea of BLOCK BENCH but emphasised the impact of workload size on the computing layer of blockchain nodes. Zheng et al. \cite{Zheng2018} proposed an alternative monitoring framework that analyses the logs generated by Ethereum clients to infer their performance in real-time. Beyond performance, researchers have also applied benchmarks to assess the correlation between CPU consumption and blockchain's rewards \cite{Aldweesh2018}, and bandwidth requirement for real-time use cases \cite{Meeuw2019}. A survey of blockchain benchmark suites is available in \cite{Wang2019}.

Simulations assess the qualities of a blockchain network before deployment. They are helpful in scenarios where deploying a blockchain network prototype is impractical. The existing simulators \cite{Yasaweerasinghelage2017,Wang2018,Zhang2018,Aoki2019} vary broadly in simulation parameters and predicted indicators. However, their inputs are primarily reflections of the architectures of blockchain networks that they need to simulate. 

NVAL does not offer new simulators or benchmarks. However, it offers a platform for integrating and employing simulators and benchmark suites. Specifically, simulators and benchmark suites can be encapsulated as composable actions and integrated into NVAL's action pool. The bridging modules of these actions can transform architectural specifications (BND fragments) into the input expected by the enclosed simulators or benchmark suites. 

\section{Conclusion}
\label{sec:Conclusion}

This paper introduces NVAL, a software framework that automatically deploys and evaluates blockchain networks based on architectural specifications. This framework enables practitioners and researchers to focus on designing blockchain networks and reduces the time and effort consumed by implementation mistakes and misconfigurations. NVAL operates by breaking down complex multi-channel blockchain networks into logical channels and leveraging the existing automation programs contributed by practitioners to deploy and evaluate these channels individually. We propose a novel meta-model to capture blockchain network architectures, a novel concept and architecture to capture automation programs, and a state-space search approach to match architectures with actions and construct execution plans automatically. A real-world case study with 65 blockchain networks featuring 12 diverse architectures shows that NVAL can accurately deploy and evaluate blockchain networks with negligible time overheads. 

We plan to extend NVAL by introducing semantic descriptions to actions and BND models to enable more sophisticated semantic-based matching between actions and architectures. We also aim to explore AI-based techniques to build actions automatically from the existing open-source deployment scripts and natural language instructions. These solutions aim to further improve the longevity and effectiveness of NVAL as an all-in-one solution for researchers and practitioners to deploy and evaluate complex blockchain networks. 


\section*{Acknowledgements}

We would like to thank Chadni, Fattah, and Shaw from the Center for Research on Engineering Software Technologies (CREST) as well as anonymous reviewers for the insightful and constructive comments.

%
\bibliographystyle{elsarticle-num} 
\bibliography{references}
\end{document}